\documentclass[pra,reprint,longbibliography,apsrev]{revtex4-1}
\usepackage[latin1]{inputenc}
\usepackage{graphicx}
\usepackage{subfigure}
\usepackage{amsmath}
\usepackage{amssymb}
\usepackage{bbold}
\usepackage{color}
\usepackage{times}
\usepackage[vcentermath]{youngtab}
\usepackage{young}
\usepackage{youngtab}
\usepackage{natbib}
\usepackage[colorlinks,bookmarks=false,citecolor=blue,linkcolor=red,urlcolor=blue]{hyperref}
\graphicspath{{Figs/}}

\DeclareMathOperator{\Tr}{Tr}
\newcommand{\SU}[1]{SU$(#1)$}

\begin{document}

\title{Structure of Spin Correlations in High Temperature \SU{N} Quantum Magnets}

\author{Christian Romen}
\author{Andreas M. L\"auchli}
\affiliation{Institut f\"ur Theoretische Physik, Universit\"at Innsbruck, A-6020 Innsbruck, Austria}
\date{\today}

\begin{abstract}
Quantum magnets with a large \SU{N} symmetry are a promising playground for the discovery of new forms of 
exotic quantum matter. Motivated by recent experimental efforts to study \SU{N} quantum magnetism in samples of ultracold
fermionic alkaline-earth-like atoms in optical lattices, we study here the temperature dependence of spin correlations
in the \SU{N} Heisenberg spin model in a wide range of temperatures. We uncover a sizeable regime in temperature, starting
at $T=\infty$ down to intermediate temperatures and for all $N\ge2$, in which the correlations have a common spatial structure on a broad range of lattices, 
with the sign of the correlations alternating from one Manhattan shell to the next, while the amplitude of the correlations is rapidly decreasing with
distance.
Focussing on the one-dimensional chain and the two-dimensional square and triangular lattice for certain $N$, we discuss the appearance of a {\em disorder} and a {\em Lifshitz}
temperature, separating the commensurate Manhattan high-$T$ regime from a low-$T$ incommensurate regime. We observe that this temperature window
is associated to an approximately $N$-independent entropy reduction from the $\ln(N)$ entropy at infinite temperature. 
Our results are based on high-temperature series arguments and as well as large-scale numerical full diagonalization results of thermodynamic 
quantities for \SU{3} and \SU{4} square lattice samples, corresponding to a total Hilbert space of up to $4\times 10^9$ states.
\end{abstract}

\maketitle

\section{Introduction}

More than a decade ago first proposals put forward the use of internal states of ultracold atoms in order 
to implement various models of \SU{N} quantum magnetism and multi-orbital physics~\cite{Honerkamp2004,Cazalilla2009,Gorshkov2010}.
In contrast to "plain" quantum simulations, where the emulation of a particular condensed matter problem is central to the effort, here the
proposals offered a new playground for theory and experiment to explore uncharted territory. The study of \SU{N} quantum magnetism started
historically as a largely academic endeavour in the context of integrable systems and large-$N$ limits of \SU{2} quantum magnetism~\cite{Sutherland1975,Affleck1988,Read1989,Marston1989},
but in the meantime a large swath of theoretical and numerical work has demonstrated that the field of \SU{N} quantum magnetism offers 
many opportunities for exciting new physics, waiting to be uncovered in experiments~\cite{Daley2011,Cazalilla2014}.

Concerning the experimental AMO platform, it turned out that the alkaline-earth(-like) fermionic atoms $^{87}$Sr~\cite{DeSalvo2010,Khoon2010,Stellmer2011,Stellmer2014,Zhang2014,Qi2019,Sonderhouse2020,Bataille2020}
and $^{173}$Yb~\cite{Fukuhara2007a,Taie2010,Sugawa2011,Pagano2014,He2020} are well suited for this line of research. On the road towards 
\SU{N} quantum magnetism of localized moments on a lattice, the realization of a Mott insulating state of $^{173}$Yb atoms formed an important milestone~\cite{Taie2012,Hofrichter2016},
paralleling the earlier achievements of \SU{2} Mott insulators~\cite{Joerdens2008,Schneider2008}. The realm of \SU{2} magnetism has seen tremendous experimental progress with the
advent of the real-space resolution of spin correlations using quantum gas microscopes and other probes~\cite{Greif2013,Parsons2016,Boll2016,Cheuk2016,Brown2017,Mazurenko2017,Hart2015}.
For $^{173}$Yb~first promising experimental results for nearest neighbor spin correlations in \SU{4} and \SU{6} Mott insulators were reported recently~\cite{Ozawa2018,TakahashiICAP2018,TakahashiDAMOP2020}, and efforts towards quantum gas microscopes for Sr or Yb atoms are on their way~\cite{Miranda2015,Miranda2017,Knottnerus2020,Okuno2020}.

The original proposals and the subsequent experimental work motivated a broad range of theoretical and computational works on various aspects
of \SU{N} quantum magnets~\cite{
Frischmuth1999,Manmana2011,Messio2012,Bonnes2012,
Yip2014,Decamp2016,Huang2019,
Nonne2013,Rachel2009,Dufour2015,Capponi2016,
Tanimoto2015,Suzuki2015,Okubo2015,Motoyama2018,Gauthe2020,
Sotnikov2014,Sotnikov2015,Yanatori2016,Golubeva2017,
KanaszNagy2017,
Assaad2005,Hazzard2012,Cai2013,Zhou2014,Wang2014,Lee2018,Tamura2019,Chung2019,Choudhury2020,
Fukushima2002,Fukushima2003,Fukushima2005,
Nataf2014,Nataf2016b}. A significant effort was put into understanding the ground state ($T=0$) phase 
diagrams of quantum spin models in the fundamental representation of \SU{N}~\cite{VanDenBossche2000,VanDenBossche2001,Penc2003,Lauchli2006,Hermele2009,Toth2010,*Toth2012,Corboz2011,Hermele2011,Corboz2012a,Corboz2012b,Bauer2012,Corboz2013,Song2013,Capponi2016,Nataf2016b,Nataf2016c,Weichselbaum2018,Keselman2019,Boos2020}. 

The experiments for Mott insulators of $^{173}$Yb~\cite{Taie2012,Hofrichter2016} operate currently in a temperature $(T)$ or entropy $(S)$ regime, which is low enough to freeze out the charge fluctuations at the repulsion energy $U$, therefore
justifying the Mott insulating regime. However the thermal entropy per particle is still substantial, so that one likely operates at effective temperatures above or around the magnetic exchange scale $J$.

In this manuscript we address the structure and temperature dependence of real-space and momentum-space spin correlations in this particular temperature or entropy regime.
We find an underlying common structure of correlations in real-space for {\em all} $N$ in \SU{N} and across many lattices. We also find that many aspects of the thermodynamics in this
regime are to a large extent $N$-independent.

\section{Model}

An appropriate starting point to describe the systems of interest is a 
single-band, \SU{N}-symmetric, fermionic Hubbard model, which models $^{173}$Yb or $^{87}$Sr atoms (in the electronic $^1S_0$ 
ground state) confined to an optical lattice~\footnote{The inclusion of the $^3P_0$ clock state leads to a multi-orbital \SU{N} Hubbard model,
which is interesting in itself~\cite{Scazza2014,Cappellini2014,Hoefer2015,Riegger2018}, but not the topic of the
present work.}.

\begin{eqnarray}
  {\cal H}_\mathrm{Hubbard}
   &=& -t\ \sum_{\langle i,j\rangle,\alpha} \Bigl(
             c^\dagger_{i,\alpha} c^{\phantom{\dagger}}_{j \alpha}
         + \mathrm{h.c.} \Bigr)\\
            && +\ \frac{U}{2} \sum_{i} n_{i} (n_{i}-1)\ . \notag
       \label{eq:Hubbard} 
\end{eqnarray}
Here $t$ denotes the tunneling amplitude for nearest neighbor bonds on the lattice
and $U>0$ parametrizes the repulsive onsite interaction strength. $c^\dagger_{i,\alpha}$ ($c^{\phantom{\dagger}}_{i \alpha}$)
are the creation (annihilation) operators of fermions with internal state $\alpha\in \{1,\ldots,N\}$ on lattice
site $i$. The operator $n_i\equiv\sum_\alpha n_{i,\alpha}=\sum_\alpha c^\dagger_{i,\alpha}$ $c^{\phantom{\dagger}}_{i \alpha}$ determines the total number of fermions on site $i$.

While the general phase diagram of this model  is to a large extent unknown, and
the charting thereof constitutes one of the goals of the experimental investigations,
it is clear that at integer fillings, in the limit of strong repulsive interactions 
$U\gg |t|$ and low temperature $T\ll U$, Mott insulating phases do occur, where charge fluctuations are suppressed, 
and the system is thus insulating, i.e. charge transport
is inhibited. In this limit the description of the system can be simplified by projecting out the local occupancies away from
the considered integer filling and therefore adopting an \SU{N} 
symmetric effective spin model. Depending on the integer filling $\langle n_i\rangle=n$, the spin model 
is formulated with local spins in the $n$-box antisymmetric irreducible representation of
\SU{N}. 
The Heisenberg spin model then reads
\begin{equation}
  {\cal H}_{\mathrm{HB}} =  J
    \sum_{\langle i,j\rangle,A} 
      S^{A}_{i}  S^{A}_{j}
\text{ ,}\label{eq:Hspin}
\end{equation}
with the antiferromagnetic coupling $J=4 t^2/U > 0$ at leading order in $t/U$~\footnote{This is the formula valid for the fundamental representation.  At higher order in $t/U$ new terms in
the spin model can be generated (see e.g.~\cite{Boos2020}), but we stick to the Heisenberg term here.}. The sum $A$
extends over the $N^2-1$ generators of \SU{N}. The dimension of the spin operators depends on the 
irreducible representation considered, and we will now focus exclusively on the case of unit filling $n=1$, corresponding to  the fundamental irreducible representation of \SU{N} (Young tableau: ${\tiny \yng(1)}$ ) of
dimension $N$. In this particular case the spin interaction can also be rewritten exactly as a sum of two-site transposition operators:
\begin{equation}
  {\cal H}_{\mathrm{HB}} =  J   \sum_{\langle i,j\rangle} 
      \tfrac{1}{2} ( P_{i j}-\tfrac{1}{N} )
\text{ ,}\label{eq:Hspin_perm}
\end{equation}
with $P_{i j} = \sum_{\alpha,\beta}
|\alpha_i,\beta_j\rangle\langle \beta_i ,\alpha_j|$, i.e.~a two site permutation operator with $\alpha,\beta \in \{1,...,N\}$. Some parts of the recent literature on
quantum spin models in the fundamental representation are working with this permutation formulation. In order to allow for
a simple comparison to the established correlations and temperature scales for \SU{2} we however continue our discussion with the 
Heisenberg Hamiltonian in the spin operator convention Eq.~\eqref{eq:Hspin}. We discuss the relation between different observables quantifying
spin correlations in App.~\ref{app:spincorrs}.

As mentioned in the introduction, our goal is to explore and characterize the structure of spin correlations in a temperature regime where
the charge fluctuations can be neglected, i.e.~at $T\ll U$. This is an experimentally relevant regime,  as some of the currently reported experiments operate 
at entropies per particle around or somewhat below $\ln N$~\cite{Hofrichter2016} in the Mott regime. In the spin language this corresponds to temperature ranges
from $T/J \sim 1$ to $T/J=\infty$. While spin correlations of one-dimensional spin chains have been studied at finite temperature in the past~\cite{Frischmuth1999,Fukushima2002,Messio2012,Bonnes2012},
there is a scarce number of works~\cite{Fukushima2005} addressing \SU{N} spin correlations at finite temperature in higher dimensions. 

Our work is based on simple high-temperature
series considerations~\cite{book_SeriesExpansionMethods}, complete numerical exact diagonalization (ED) of periodic finite size clusters using a basis of \SU{N} Young tableaux~\cite{Nataf2014} and 
numerical linked cluster expansion~\cite{Rigol2007a} results (also in the \SU{N} Young tableaux basis), and aims to explore the structure and temperature 
behaviour of spin correlations in the currently experimentally accessible temperature or entropy regime in \SU{N} Mott insulators across a variety of (mostly two-dimensional) lattices. 

\begin{figure}
\centering
\includegraphics[width=\linewidth]{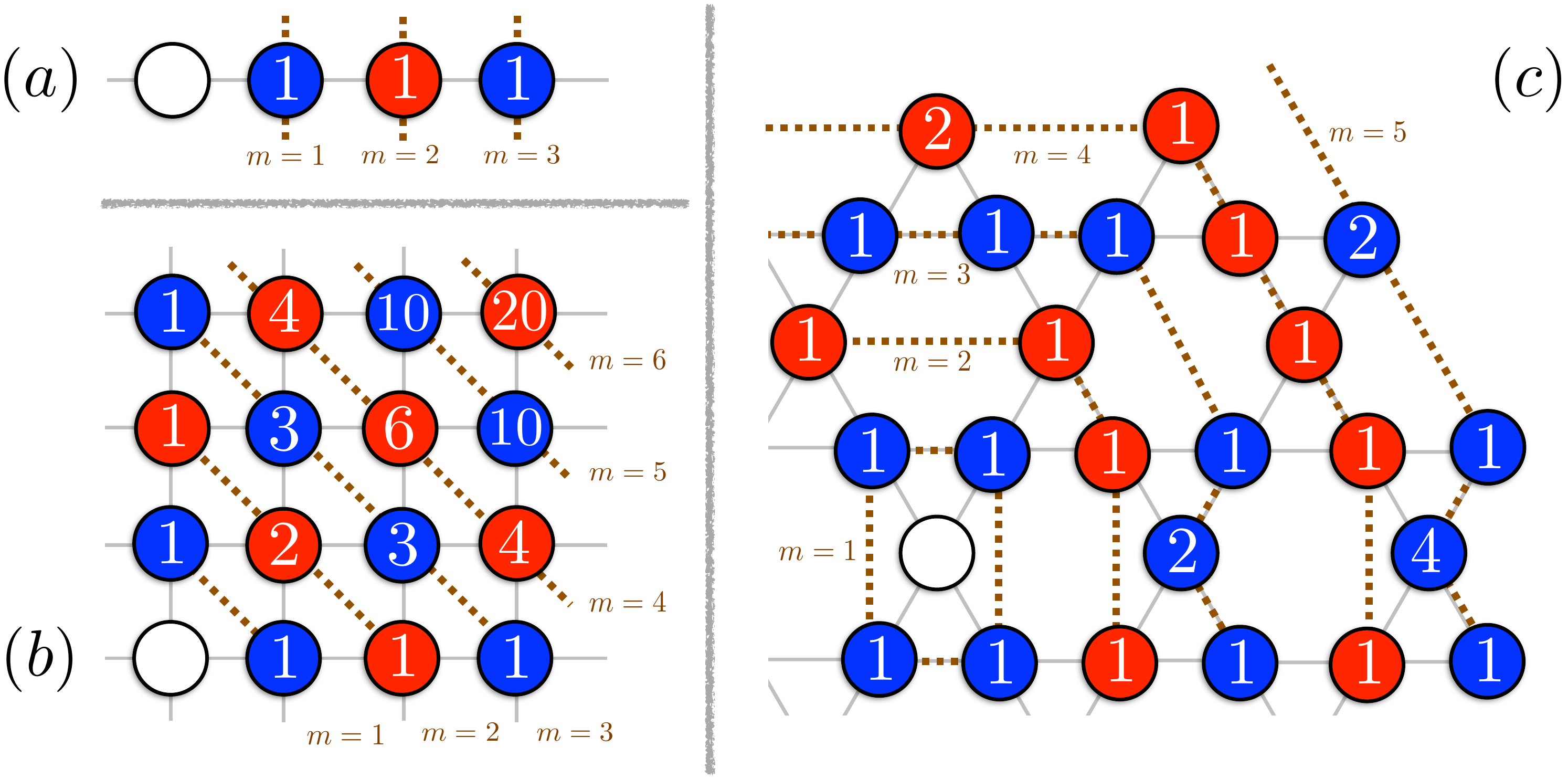}
\caption{Manhattan physics for \SU{N} magnets. Qualitative real-space two-site correlation pattern in the high-temperature regime  $J {\ll} T {\ll} U$ of the \SU{N} Heisenberg model on different geometries:
(a) one dimensional chain, (b) square lattice and (c) kagome lattice. 
Blue (red) denotes a negative (positive) correlation with the reference site (empty circle symbol).
 The sign of the correlator only depends on the Manhattan distance $m$ to the reference site.
The dotted brown line connects equidistant sites from the reference site and forms Manhattan-shells denoted by $m$,
i.e.~nearest neighbors to the reference site are found in the first shell $m=1$, then at distance two $m=2$ the second shell, etc..
The prefactor $g_\mathrm{paths}(\mathbf{r})$ of the corresponding correlator is indicated at the lattice sites. $g_\mathrm{paths}(\mathbf{r})$ 
counts the number of possible paths connecting the two sites with $m$ Manhattan distance steps (using only nearest neighbor bonds).}
\label{fig:manhattan_overview}
\end{figure}

\section{Spin Correlations at high Temperature}

We start with a simple, but insightful, high-temperature series consideration. We are interested 
in the following temperature dependent spin correlations at distance 
$\mathbf{r}\neq \mathbf{0}$~\footnote{The autocorrelation is independent of temperature: $\sum_{A} S^{A}_{\mathbf{0}}  S^{A}_{\mathbf{0}}=(N^2-1)/(2N)$}:
\begin{equation}
C_\mathrm{spin}(T,\mathbf{r})= \Tr \left[\hat{\rho}(T)\ \sum_{A} S^{A}_{\mathbf{0}}  S^{A}_{\mathbf{r}}\right]
\end{equation}
The density matrix 
\begin{equation}
\hat{\rho}(T)=\frac{\exp(-\mathcal{H}_\mathrm{HB}/T)}{Z(T)}
\end{equation}
is the standard normalized canonical Gibbs density matrix (we set $k_B=1$), where $Z(T)=\Tr[\exp(-\mathcal{H}_\mathrm{HB}/T)]$ denotes the partition function. We assume a translation invariant situation
and choose the reference site at the origin $\mathbf{0}$. For other observables suitable to capture \SU{N} spin correlations and their relation, please
refer to App.~\ref{app:spincorrs}.

At infinite temperature $(T/J=\infty)$ in the spin model \eqref{eq:Hspin} all spin correlations
as defined above vanish (c.f.~App.~\ref{app:spincorrs}), irrespective of the value of $N$.

Let us now discuss the leading order behaviour in $\beta J = J/T$ for a correlation at distance $\mathbf{r}$
for a general lattice made of nearest neighbour bonds of the same strength. Standard linked cluster arguments
for high-temperature series~\cite{book_SeriesExpansionMethods} imply that the correlator at distance $\mathbf{r}$
starts at an order $k$ in $(\beta J)$ which is directly linked to the Manhattan distance $m$ between the origin $\mathbf{0}$ and 
site $\mathbf{r}$. In Fig.~\ref{fig:manhattan_overview} we indicate the Manhattan distance $m$ between the reference site
and a few shells of sites on (a) the linear chain, (b) the square lattice and (c) the kagome lattice for illustration. The structure on other
lattices can be derived accordingly. As discussed below an additional element of the leading
order expression concerns the number of shortest paths $g_\mathrm{paths}(\mathbf{r})$, measured in the Manhattan metric, 
which link the origin to the site of interest. These numbers $g_\mathrm{paths}(\mathbf{r})$ are marked in the site circles 
in Fig.~\ref{fig:manhattan_overview}.

The considerations so far are independent of the actual Hamiltonian, as long as it consists only of nearest neighbor bonds on the lattice.
There are also interesting connections to the short-time expansion of correlations when starting from an uncorrelated product state, as 
recently discussed and experimentally demonstrated in a Rydberg quantum magnetism experiment on square and the triangular lattices~\cite{Lienhard2018}.

For our Hamiltonian at hand, when written in the permutation formulation~\eqref{eq:Hspin_perm}, it is possible to explicitly calculate the coefficient
of the leading order high-temperature expression symbolically for all $N$~\cite{Fukushima2002,Fukushima2003} for the first few orders $k\le 12$. 
Given the simple structure of the terms we conjecture that the expression holds for all $k$. Based on these
results we are now in a position to present the leading order high-temperature expression for general $N$ on any lattice in any dimension, as long as the
Hamiltonian~\eqref{eq:Hspin} only contains nearest-neighour bonds of equal strength:
\begin{eqnarray}
  C_\mathrm{spin}(T,\mathbf{r})  &=& \frac{g_\mathrm{paths}(\mathbf{r})}{2} (-K)^m \left( \frac{1}{N^{m-1}} - \frac{1}{N^{m+1}} \right)\nonumber\\
  		&&+O(K^{m+1}),
		\label{eq:highT_spincorrs}
\end{eqnarray}
where $K=\beta J/2$, and $m$ is the Manhattan distance between sites $\mathbf{0}$ and $\mathbf{r}$ on the considered lattice,
while $g_\mathrm{paths}(\mathbf{r})$ counts the number paths between the two sites of length $m$, measured in the Manhattan metric.

This is a central result of our work. The expression~\eqref{eq:highT_spincorrs} allows us to predict both
the real-space structure of the correlations, and their relative $N$ dependence in the high-temperature regime of Hamiltonian \eqref{eq:Hspin}, within the limits of its applicability, to be discussed below.

As one can see from the term $(-K)^m$, the spin correlations are alternating in sign from one Manhattan shell to the next, starting with the nearest-neighbour correlations being negative, i.e.~antiferromagnetic,
as expected for an antiferromagnetic spin Hamiltonian. This feature has been noticed before in earlier work on linear chains~\cite{Fukushima2002} and the cubic lattice~\cite{Fukushima2005}.
Furthermore sites within the same Manhattan shell $m$ are more correlated by a factor $g_\mathrm{paths}(\mathbf{r})$, when multiple paths link the 
two considered sites. These enhancement factors are displayed in the circles in Fig.~\ref{fig:manhattan_overview}. 

The dependence on $N$ in \SU{N} is also remarkable. For $m=1$, i.e.~nearest neighbours, the correlations are proportional to $(1-1/N^2)$, indicating that the correlations
are actually {\em increasing} with $N$ and converging to a constant value at large $N$ (for a given $K$). This is also an important feature, and its thermodynamic implications 
will be discussed further in Sec.~\ref{sec:eqationofstate}. For more distant sites with $m>1$ the correlations are proportional to $(1/N^{m-1}-1/N^{m+1})$
indicating that the correlations {\em decrease} rapidly in magnitude with $m$ and converge towards zero as $N$ grows, and even more strongly so for larger $m$. 

\begin{figure}[!htb]
\centering
\includegraphics[width=0.9\linewidth]{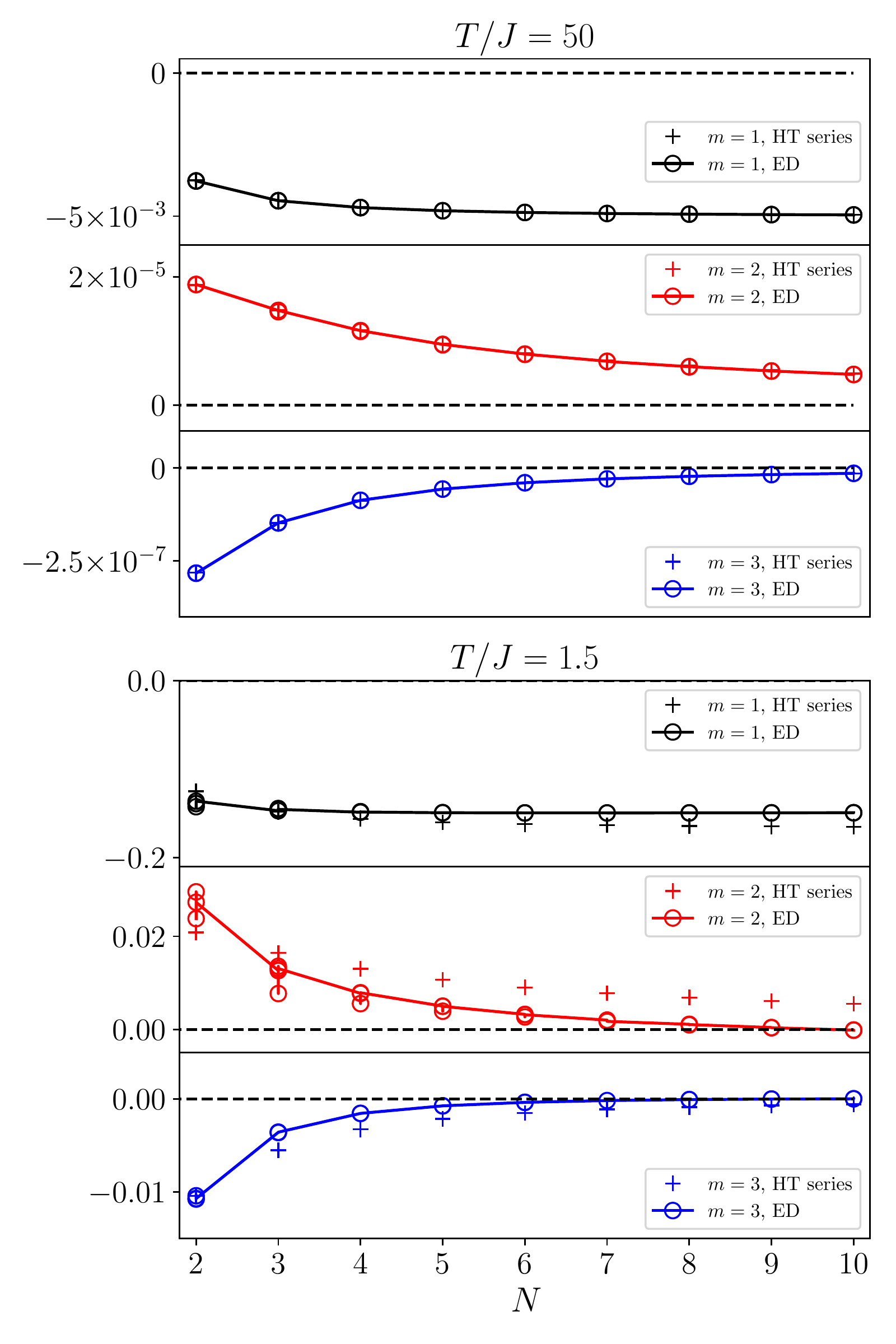}
\caption{Real space spin correlations on the square lattice for two different temperatures: $T/J=50$ for
the upper panel (with distances $m=1,2,3$) and $T/J=1.5$ for the lower panel (for the same distances). 
The empty circles denote numerical exact diagonalization results while the crosses correspond to the leading order high-temperature
series result presented in Eq.~\eqref{eq:highT_spincorrs}.
\label{fig:real_space_corrs}}
\end{figure}

As a first step towards addressing the range of applicability of the leading-order high temperature series argument just developed,
we display in Fig.~\ref{fig:real_space_corrs} numerical exact diagonalization results for several finite-size square 
lattice clusters at different temperatures $T/J$ and distances $m$ as a function of $N\in\{2,\ldots,10\}$~\footnote{Where applicable we have divided the numerical results by 
$g_\mathrm{paths}(\mathbf{r})$, in order to simplify the presentation. The ED cluster have periodic boundary conditions and range in size $N_s$ from 12
to 18, depending on the value of $N$.}. In  the upper three panels of Fig.~\ref{fig:real_space_corrs}
we start at a rather high temperature of $T/J=50$. We show three panels, one for each $m=1$ (black), $2$ (red), and $3$ (blue). At this
temperature, for the distances shown, the agreement between the high-temperature series expansion result~\eqref{eq:highT_spincorrs} (crosses) and the ED results (circles)
is very good. One can also clearly see the rapid decay of the correlations with the distance $m$ and, for $m>1$, with $N$. The values of the correlations themselves are extremely 
small at this temperature. For the substantially lower
temperature $T/J=1.5$ the lower three panels highlight how the quantitative agreement between the result~\eqref{eq:highT_spincorrs} and the numerics starts to deteriorate, however
the qualitative trends remain unaltered and even a semi-quantitative agreement is visible. These numerical results provide strong evidence that the Manhattan picture introduced here prevails for a sizeable range in temperature and spatial extent for a broad range of values of $N$.

\begin{figure*}[!htb]
\centering
\subfigure[]{\includegraphics[width=0.4\linewidth]{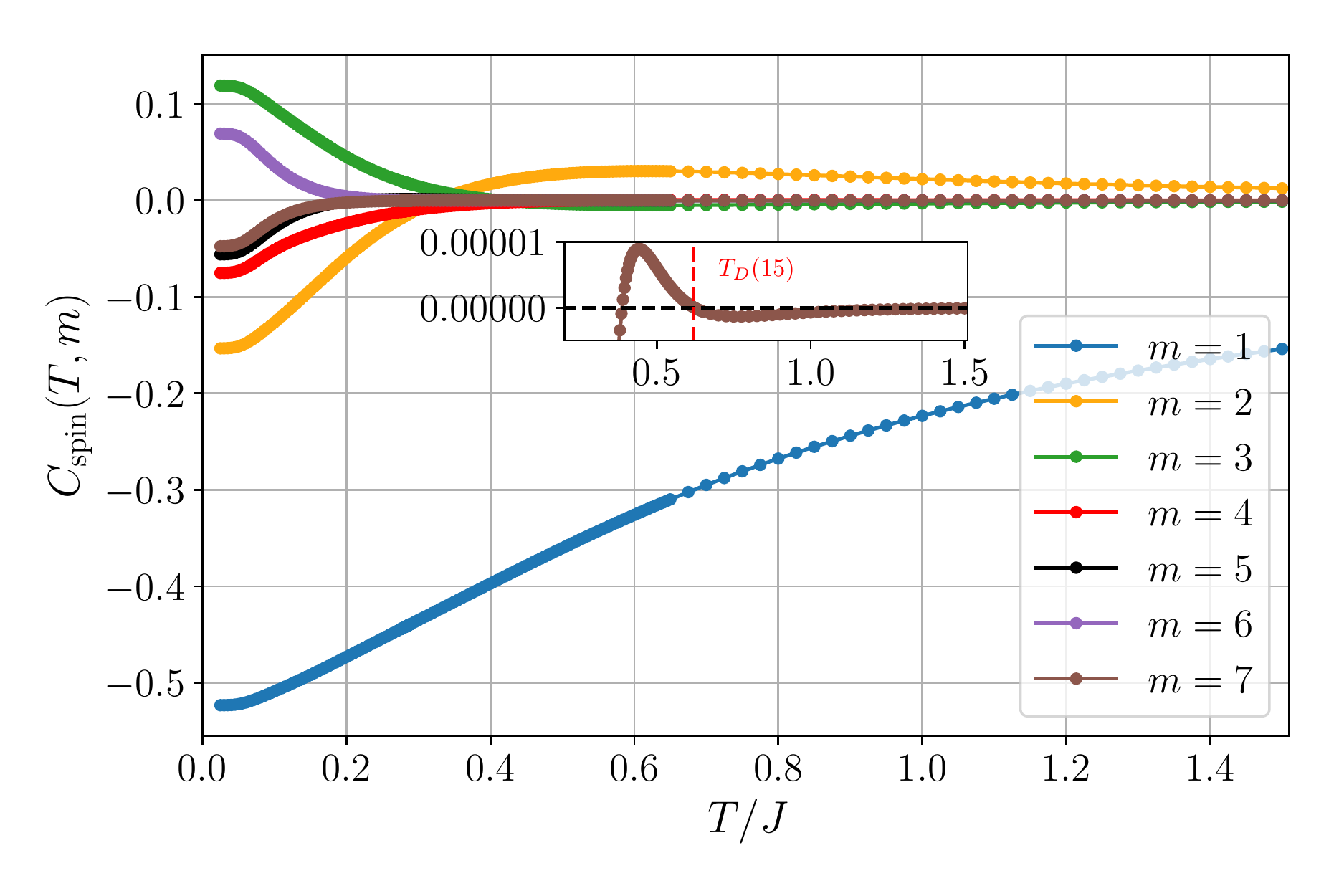}}
\subfigure[]{\includegraphics[width=0.4\linewidth]{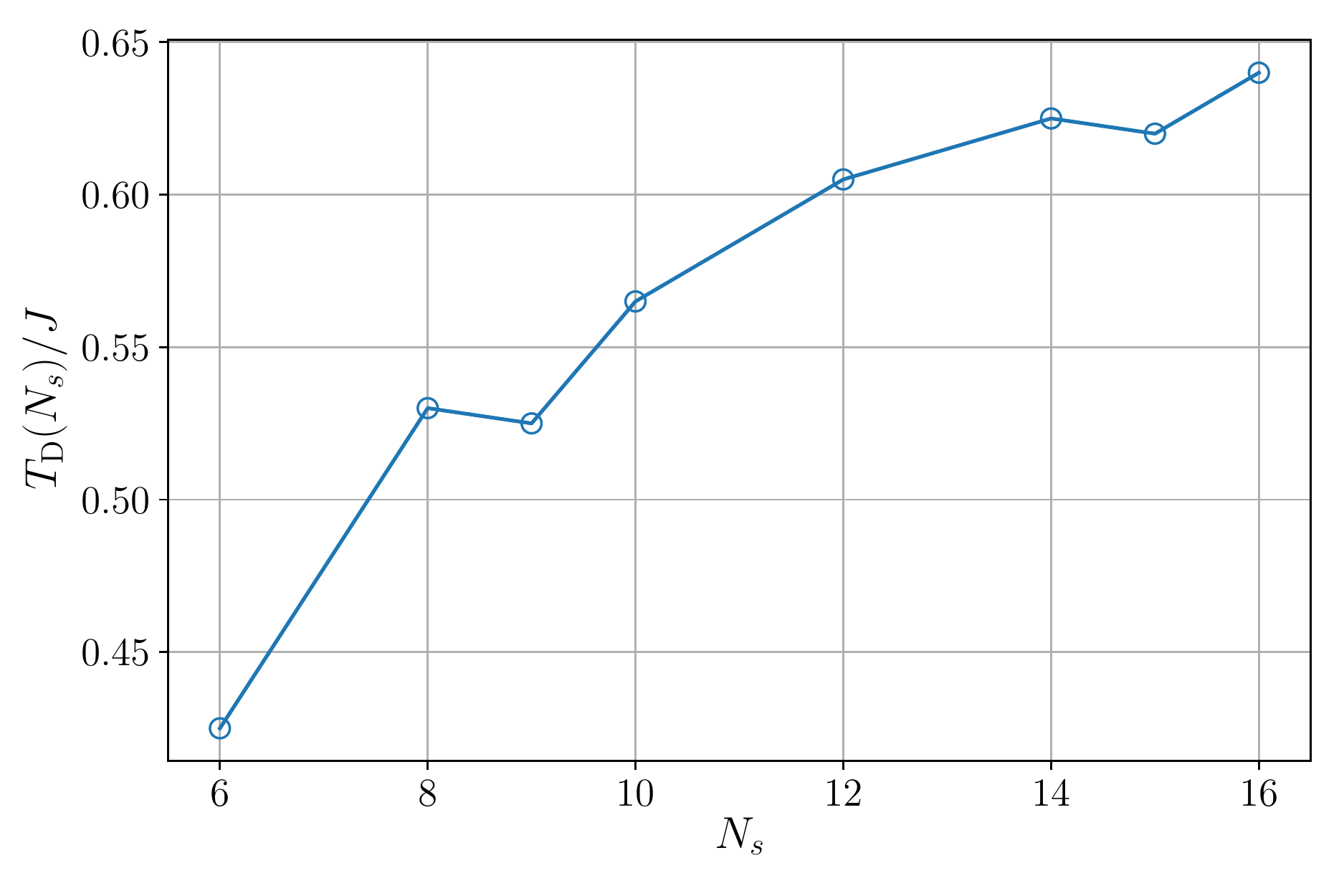}} 
\subfigure[]{\includegraphics[width=0.4\linewidth]{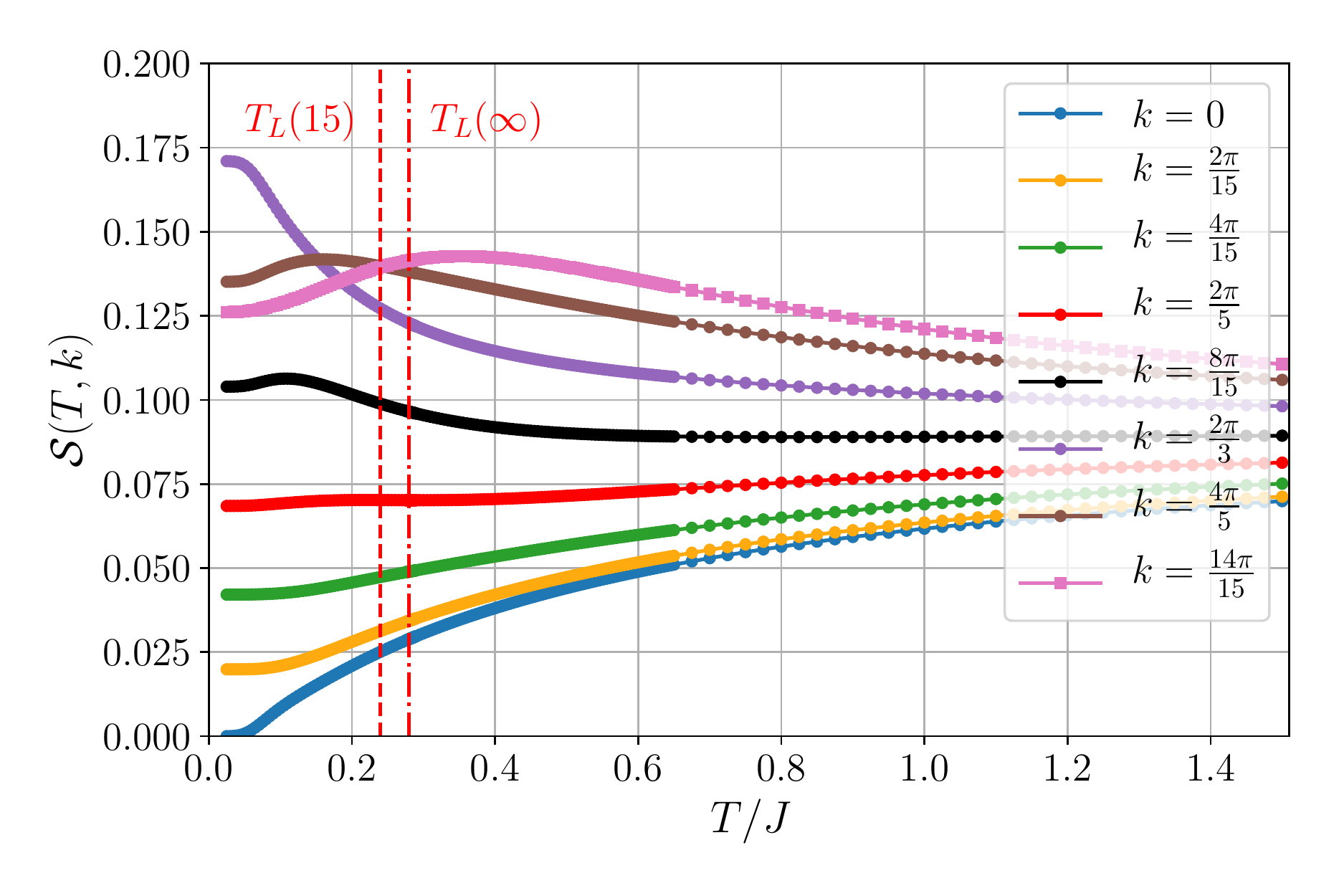}}
\subfigure[]{\includegraphics[width=0.4\linewidth]{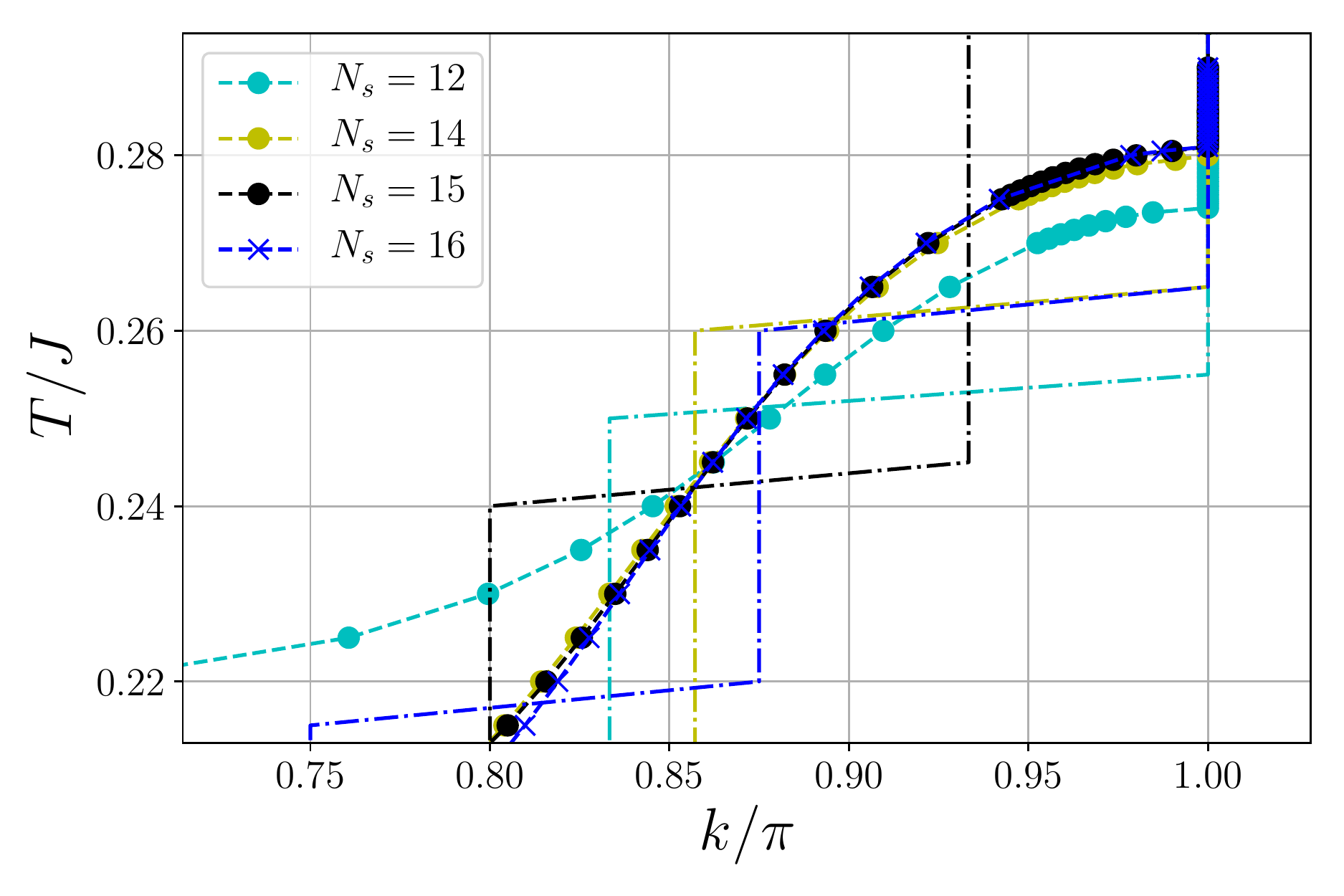}}
\caption{{\bf Temperature dependence of spin correlations in the one-dimensional \SU{3} Heisenberg model:}
(a) Real-space spin-spin correlations $C_\mathrm{spin}(T,m)$  between sites at distance $m$ as a function of temperature $T/J$ for a $N_s=15$ sites chain.
The inset resolves the sign change occurring as a function of $T/J$ for the $m=7$ correlator.
(b) Disorder temperature $T_D(N_s)/J$ as a function of system size $N_s$. 
(c) Static structure factor as a function of temperature $T/J$. The vertical red dotted and dashed dotted lines indicate the finite size bifurcation temperature $T_L(N_s=15)/J\approx0.24$ 
 (on the periodic $15$-site chain) and the estimated infinite system size Lifshitz temperature $T_L(\infty)/J=0.281(5)$.
 (d) Evolution of the peak location $k/\pi$ of the structure factor as a function of temperature $T/J$ for different chain lengths $N_s$. Dashed dotted lines are finite-size ED data and display jumps when the peaks transition from one finite size momentum to another, using a temperature grid of $\Delta T/J = 0.005$. Solid lines with symbols correspond to results obtained using the continuous structure factor ansatz of App.~\ref{app:SofKAnsatz}.
\label{fig:1d_su3_correlations}}
\end{figure*}

\section{Disorder temperatures and Lifshitz transitions}

In this section we study in more detail for which distances and temperatures the Manhattan picture starts
to break down. The ground state physics of the spin Hamiltonian \eqref{eq:Hspin} has been explored for many values of 
$N$ and lattice geometries over the last decades, see e.g.~Refs.~\cite{Sutherland1975,VanDenBossche2000,VanDenBossche2001,Penc2003,Lauchli2006,Hermele2009,Toth2010,*Toth2012,Corboz2011,Hermele2011,Corboz2012a,Corboz2012b,Bauer2012,Corboz2013,Song2013,Capponi2016,Nataf2016b,Nataf2016c,Weichselbaum2018,Keselman2019,Boos2020}, and in most cases the structure of the ground states differs qualitatively from
the Manhattan picture advocated in the previous section. Based on the current understanding, we expect only bipartite lattices with $N=2$ to show a common 
sign structure of correlations from high to low temperatures. In all other cases we have to assume the Manhattan picture to break down at some temperature in one way or the 
other. In the following we discuss some scenarios on how this breakdown might occur.

We start a few general considerations and discuss the notions of a {\em disorder} temperature, a {\em Lifshitz} temperature, and a thermal first order
phase transition. We then apply these notions to discuss the well understood one-dimensional chain case for $N=3$ first, and then switch
to two-dimensional square lattices, where we focus on $N=3$ and $N=4$. We close this section with a brief analysis of the $N=3$ model on the triangular lattice.

\subsubsection{Disorder temperature $T_D(N_s)$}

For each distance, the high-temperature expansion starts with the expression given by Eq.~\eqref{eq:highT_spincorrs}. We have however no
detailed understanding of the form of the coefficient of the next order contribution in the high-temperature expansion. Such an analysis would require
a fully fledged series expansion machinery as in Refs.~\cite{Fukushima2002,Fukushima2003,Fukushima2005}, which is however not the goal of the present work. In order to quantify the deviation
we follow an idea put forward in the context of commensurate-incommensurate transitions of short-range ordered magnetic systems~\cite{Stephenson1969,Stephenson1970,Schollwock1996}. 
In such systems, the transition from a commensurate regime to an incommensurate regime can be detected in real-space or momentum space. In real
space, one diagnostic is to determine the transition point (as a function of a parameter, such as a coupling in the Hamiltonian, or in our case the temperature $T/J$)
by locating the first sign change in a correlator at any distance deviating from the commensurate structure. In our case the commensurate region is the one with the
alternating Manhattan shell structure. The parameter location is called a {\em disorder} point. Since in our case at hand we are interested in the temperature dependence, we
call this system size dependent temperature, the {\em disorder} temperature $T_D(N_s)$. Note that this temperature does not necessarily indicate a thermodynamic phase transition,
just a change in the nature of short range correlations.

\subsubsection{Lifshitz temperature $T_L(N_s)$}

Another diagnostic to track the change from commensurate to incommensurate behaviour is to determine when the peak in the corresponding structure factor is moving away from
a commensurate location. The structure factor is defined as the Fourier transform of the real-space correlations:
 \begin{equation}
 \mathcal{S}(T,\mathbf{k}) = \frac{N^2-1}{2N} + \sum_{\mathbf{r}\neq \mathbf{0}} \left(C_\mathrm{spin}(T,\mathbf{r})\ \exp[{i\mathbf{k}\cdot \mathbf{r}}]\right) .
 \end{equation}
In two of the specific geometries discussed below, the linear chain and the square lattice, the Manhattan shell structure of correlations in real space leads to a peak
in the structure factor at momentum $\pi$ or $(\pi,\pi$) respectively. We can then track the structure factor as a function of temperature $T/J$ and detect the first temperature, coming from $T/J=\infty$,
where the location of the maximum starts to deviate {\em continuously} from $\pi$ or $(\pi,\pi)$, as in a bifurcation transition. This (possibly finite-size dependent) temperature is called the Lifshitz temperature $T_L(N_s)$. In analogy to
the disorder temperature, the Lifshitz temperature is not necessarily an indication for a thermodynamic phase transition.

\subsubsection{First order phase transition}

In dimensions higher than one, another distinct possibility is that the Manhattan regime is separated from one or several low-temperature regimes by a genuine thermal {\em first order} phase transition.
In such a scenario correlations in real space would change discontinously for many distances, and the structure factor location is also expected to jump discontinously away from the commensurate position. 

\begin{figure*}[!htb]
\centering
\subfigure[]{\includegraphics[width=0.4\textwidth]{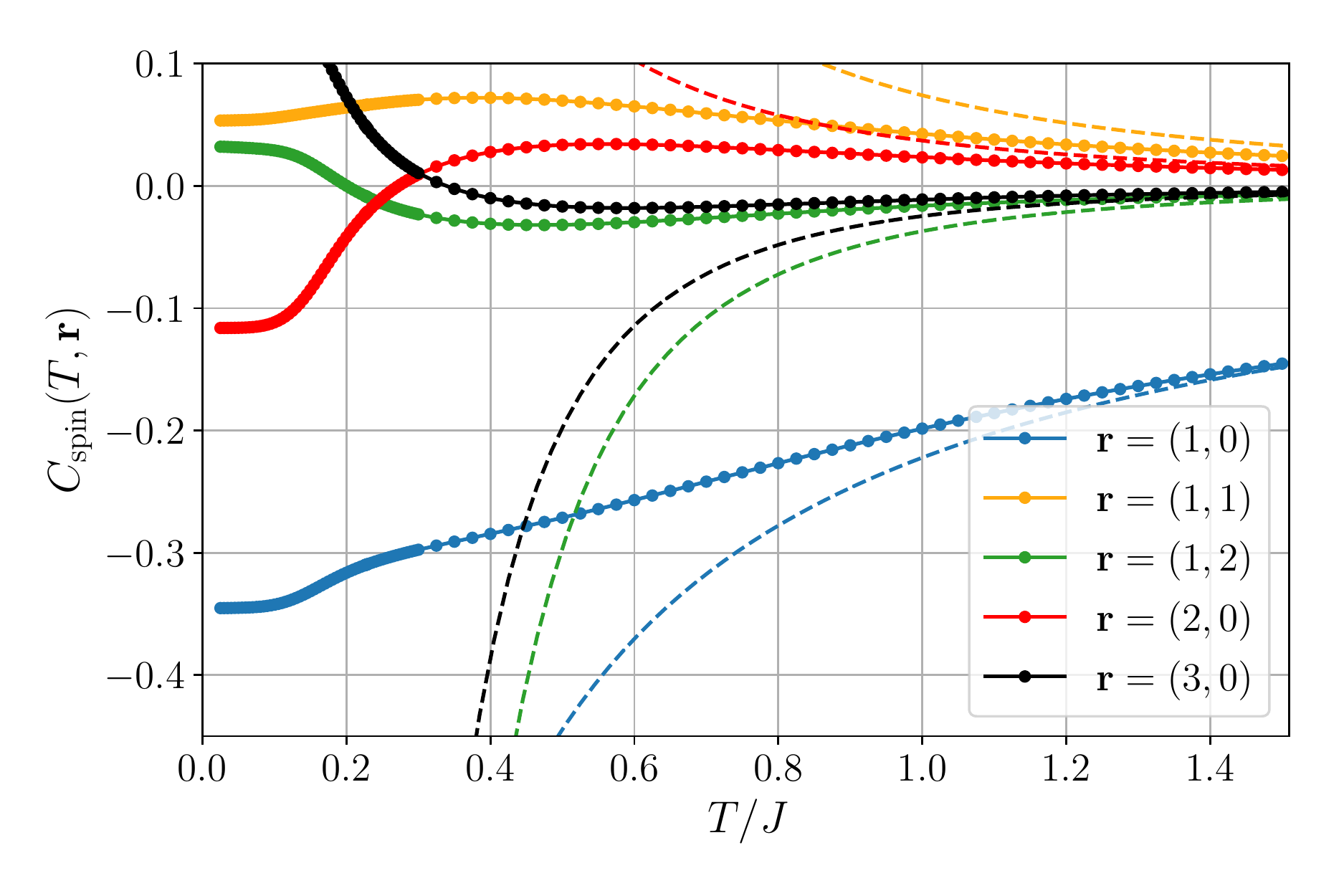}}
\subfigure[]{\includegraphics[width=0.4\textwidth]{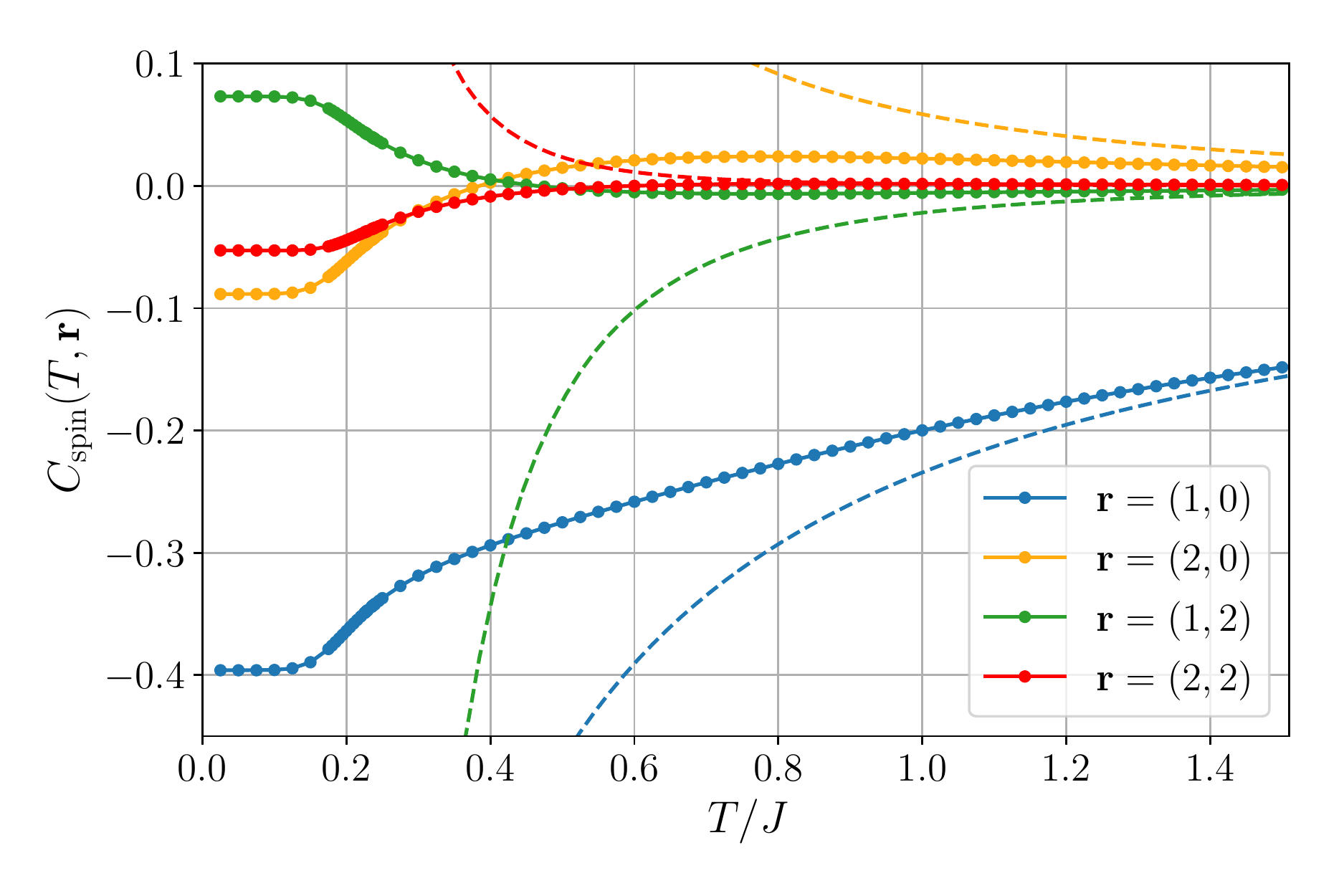}}
\subfigure[]{\includegraphics[width=0.4\textwidth]{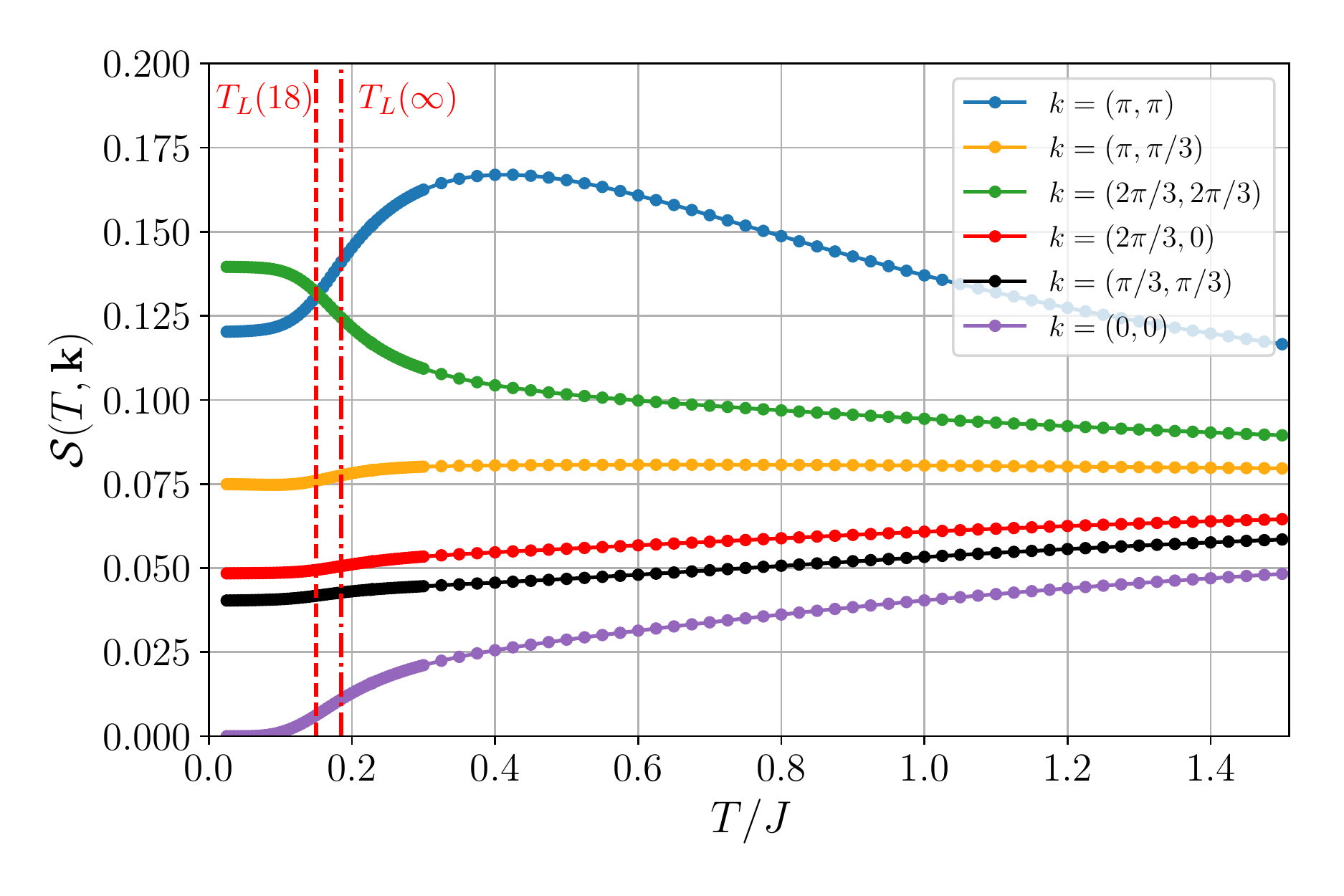}}
\subfigure[]{\includegraphics[width=0.4\textwidth]{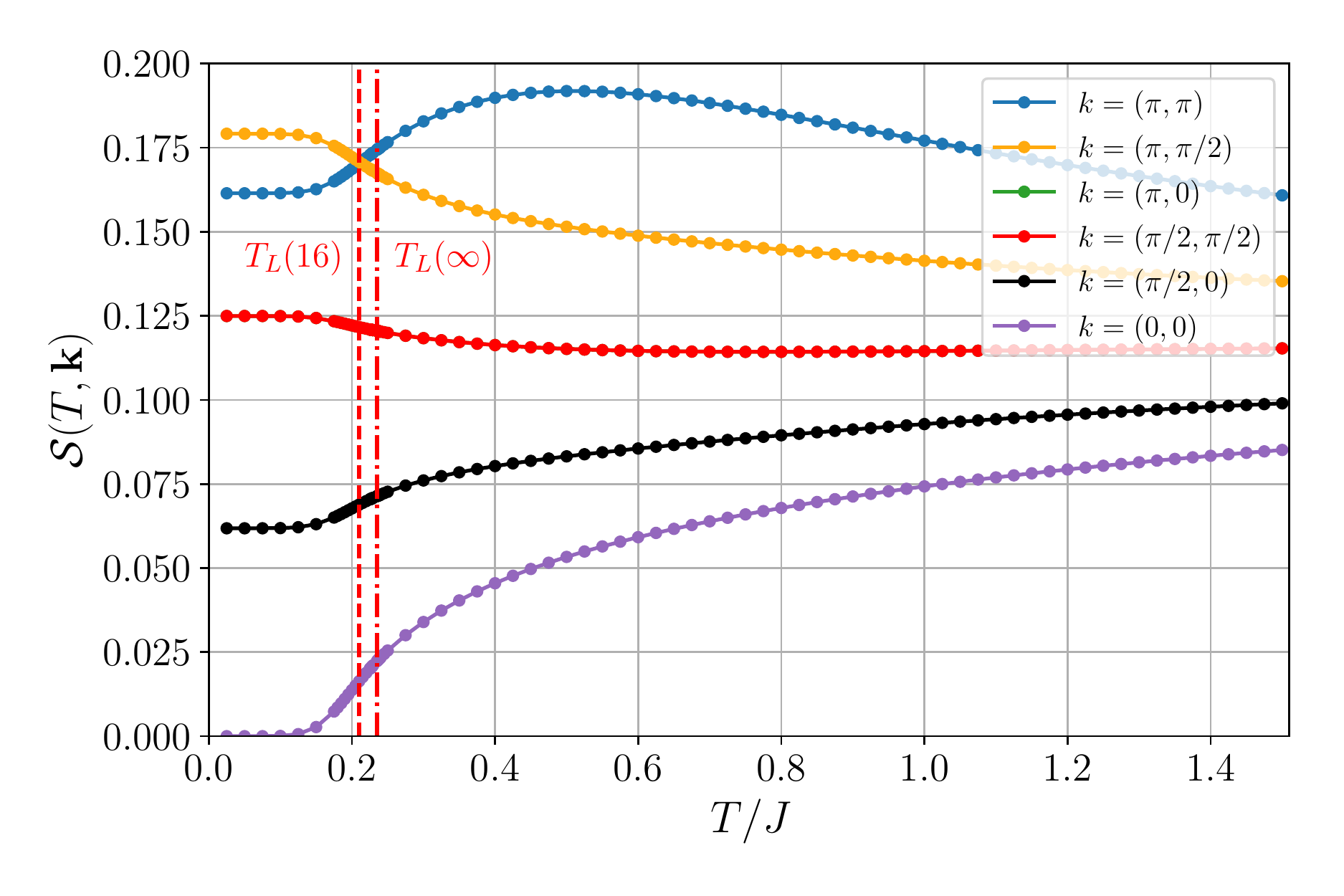}}
\caption{{\bf Temperature dependence of spin correlations in the two-dimensional square lattice \SU{3} and \SU{4} Heisenberg model:} 
Full ED spin correlation results for the largest accessible clusters for SU($3$) ($N_s=18$) and SU($4$) ($N_s=16$). 
    (a) and (b) shows the correlation functions $C_\mathrm{spin}(T,\mathbf{r})$ with a fixed reference site $0$ for $N=3$, $N_s=18$ and $N=4$, $N_s=16$, respectively. 
 Leading order high-temperature expansion results [Eq.~\eqref{eq:highT_spincorrs}] are illustrated as dashed lines.
 (c) and (d) shows 
    the corresponding structure factor $\mathcal{S}(T,\mathbf{k})$. The vertical red dashed and dashed dotted lines indicate the finite size bifurcation temperature $T_L(N_s{=}18)/J=0.150(5)$ [$T_L(N_s{=}16)/J=0.210(5)$] and
    the estimated infinite system size Lifshitz temperature $T_L(\infty)/J=0.180(5)$ [$T_L(\infty)/J=0.230(5)$ for \SU{4}], obtained by the continuous structure factor ansatz.}
\label{fig_su3_su4_correl}
\end{figure*}

\subsection{One-dimensional $N=3$ chain}

As a warm-up application of these notions we discuss the \SU{3} spin chain. The disorder temperature has not been discussed yet for \SU{N} Heisenberg chains, to the best of our knowledge. 
The Lifshitz temperature has been discussed under a different name in Ref.~\cite{Fukushima2002}.

For all $N$ the Manhattan regime of one-dimensional chains is characterized by alternating correlations as 
shown in Fig.~\ref{fig:manhattan_overview}(a) and a maximum in the structure factor at wave vector $k=\pi$. We proceed by analyzing finite-size complete ED results for \SU{3} chains up to $N_s=15$.
For each finite system size at high enough temperature the sign (and approximate values) of the correlations are
given by Eq.~\eqref{eq:highT_spincorrs}. In Fig.~\ref{fig:1d_su3_correlations}(a) we display the real-space correlations as a function of the temperature $T/J$ for a system size $N_s=15$. 
For high temperatures all real-space correlators indeed exhibit the sign predicted by the Manhattan regime, i.e.~the correlations alternate from one site to the next.
However at $T /J = T_D(N_s=15)/J\approx 0.62 J$ the first correlator changes its sign, see the inset for $m=7$ in Fig.~\ref{fig:1d_su3_correlations}(a).
At even lower temperatures other correlators change their sign, e.g.~at distance $m=2$ the correlation changes sign around $T/J\approx 0.3$. In Fig.~\ref{fig:1d_su3_correlations}(b) we plot 
the system size dependence of  $T_D(N_s)/J$ for the \SU{3} linear chain. We observe a substantial drift of these disorder temperatures towards higher values as $N_s$ increases~\footnote{We also observe some small modulation in $N_s$ with a period three.}. It is not clear to us whether this disorder temperature will drift to infinite temperature as the system size increases, or wether it will converge to a finite value $T_D(\infty)$.
Assuming the scenario of a finite disorder temperature, a linear extrapolation in $1/N_s$ yields $T_D(\infty)/J \approx 0.74(2) $.
Irrespective of this uncertainty, we interpret our observations as an indication that the real-space extent and the temperature extent of our proposed Manhattan structured correlation regime is substantial enough that
it will be able to be explored in near-term experiments measuring spin correlations beyond nearest neighbour distances.

Next we consider the correlations in momentum space by investigating the structure factor $\mathcal{S}(T,\mathbf{k})$ of the \SU{3} Heisenberg chain. 
At infinite temperature the structure factor is flat throughout the Brillouin zone. At high but finite temperature the structure factor shows
a broad peak at momentum $k=\pi$ for even $N_s$ or at $k=\pi\pm \pi/N_s$ for odd $N_s$~(see also Ref.~\cite{Fukushima2002,Bonnes2012,Messio2012}). 
This is shown in Fig.~\ref{fig:1d_su3_correlations}(c). For $N_s=15$
we see a shift of the location of the maximum from $k=14\pi /15$ to $k=4\pi /5$ at the finite size Lifshitz temperature $T_L(15)/J\approx0.24$. This is followed by a
further change from $k=4\pi /5$ to $k=2\pi /3$ around $T/J\approx 0.15$. The location of the low-temperature peak is in agreement with the known ground state physics
of \SU{N} Heisenberg chains, which display algebraically decaying spin correlations oscillating with wave vectors which are multiples of $|k|=2\pi/N$~\cite{Sutherland1975,Fukushima2002,Bonnes2012,Messio2012}. In
Fig.~\ref{fig:1d_su3_correlations}(d) we analyze the finite size dependence of Lifshitz temperature $T_L(N_s)$ using an expansion of the structure factor around $k=\pi$
discussed in App.~\ref{app:SofKAnsatz}. The analysis leads to a Lifshitz temperature of $T_L(\infty)/J=0.281(5)$ for the one-dimensional \SU{3} Heisenberg chain. In Ref.~\cite{Fukushima2002}
the Lifshitz temperature was discussed for $N=4$ and $N=20$ and a Lifshitz temperature of about $T/J\approx 0.25$ in our units was found, lending support for an approximatively  constant
Lifshitz temperature for $N\ge 3$. In Ref.~\cite{Messio2012} it was noticed on the other hand that the entropy per spin corresponding to these Lifshitz temperatures is increasing with $N$. We
will come back to this observation in Sec.~\ref{sec:eqationofstate}.

Let us  summarize for the $N=3$ Heisenberg chain that coming from high temperature, the first signal in temperature is likely the disorder temperature $T_D/J \lesssim 0.73$ (depending on distance or system size), 
where the sign structure in real space starts to show defects with respect to the Manhattan structure. At a lower temperature $T_L/J\sim 0.28$ the peak in the structure factor starts to move away from the
commensurate $\pi$ location. So real-space correlations are indeed a valuable new probe to investigate \SU{N} magnetism, as they are able to detect deviations from the Manhattan regime at higher $T/J$ than
momentum space probes.

\begin{figure*}[!t]
\centering
\subfigure[]{\includegraphics[width=0.4\linewidth]{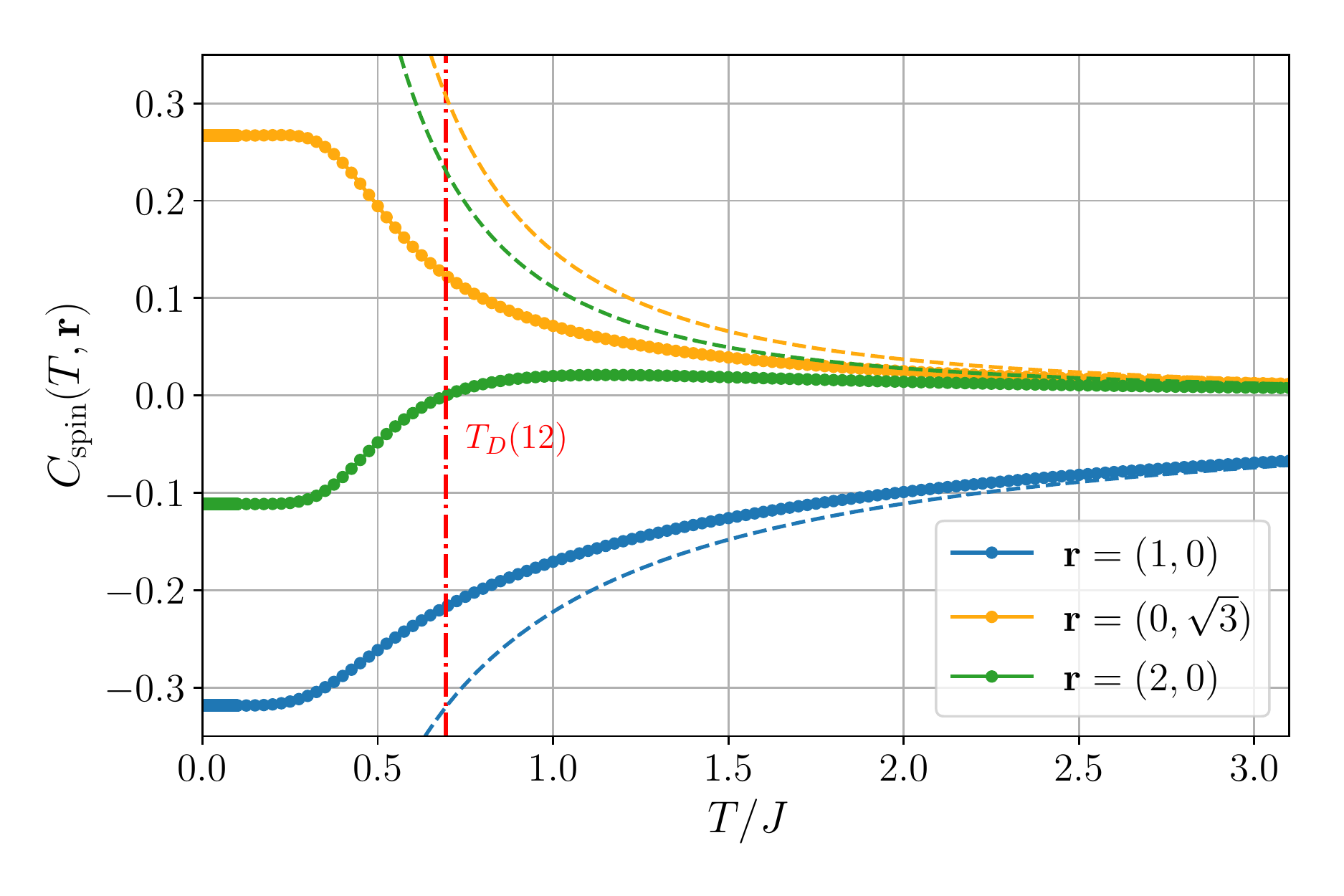}}
\subfigure[]{\includegraphics[width=0.4\linewidth]{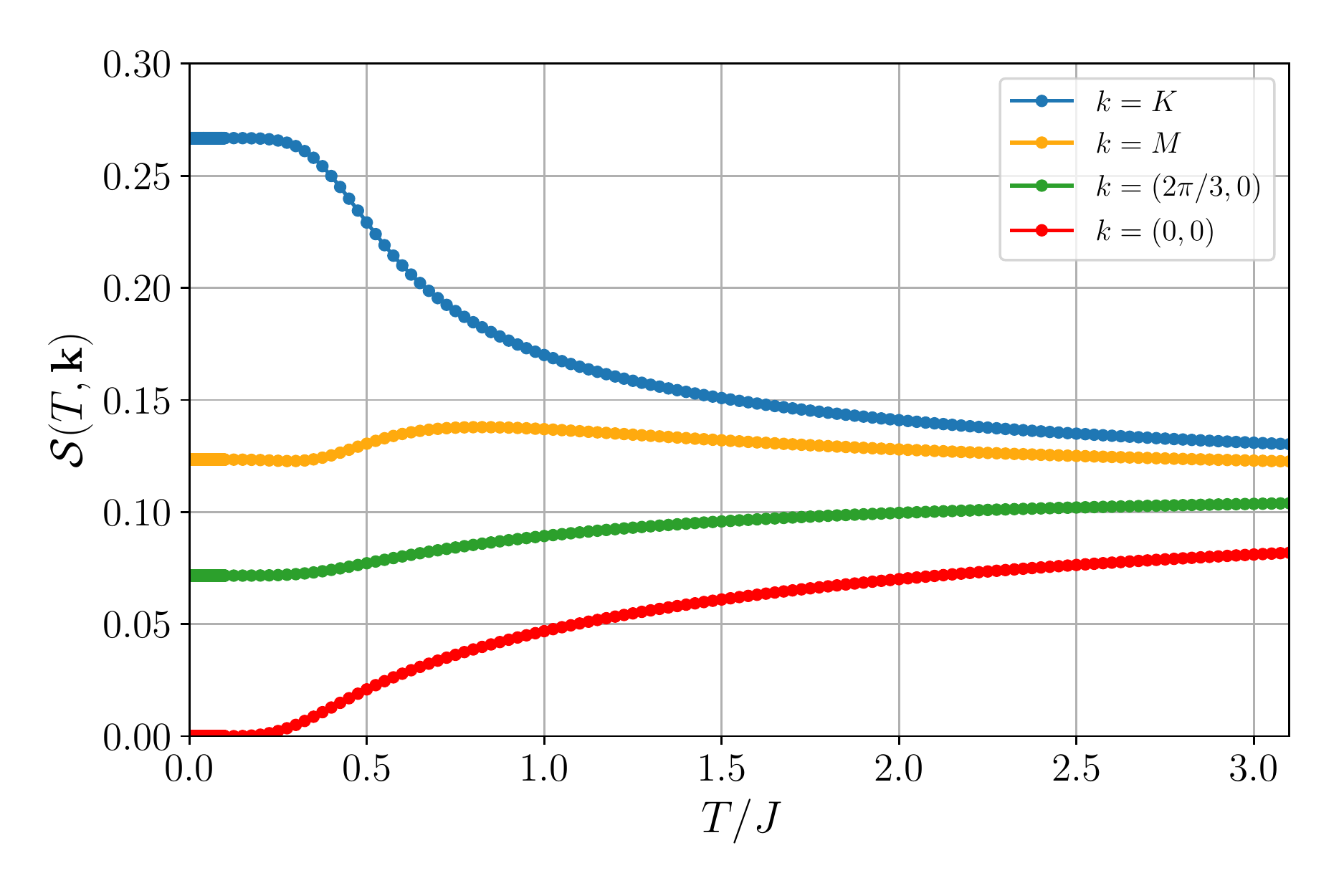}} 
\caption{{\bf Temperature dependence of spin correlations in the two-dimensional triangular lattice \SU{3} Heisenberg model:} 
Full ED spin correlation results for $N_s=12$. 
    (a) shows the correlation functions $C_\mathrm{spin}(T,\mathbf{r})$ with a fixed reference site $0$. 
 Leading order high-temperature expansion results [Eq.~\eqref{eq:highT_spincorrs}] are illustrated as dashed lines.
    The vertical red dashed dotted line indicates a disorder temperature $T_D(N_s=12)/J \approx 0.69$, where the distance $2$ correlator changes sign.
 (b) shows the corresponding structure factor $\mathcal{S}(T,\mathbf{k})$. 
\label{fig:2d_triangular_su3_correlations}}
\end{figure*}

\subsection{Square Lattices: $N=3$ and $N=4$}
\label{subsec:square_N_3_4}

The Manhattan regime for the square lattice for all $N$ exhibits real-space correlations according to Fig.~\ref{fig:manhattan_overview}(b), while in momentum space
the structure factor peaks at $(\pi,\pi)$. On the other hand the predicted ground state physics scenarios differ starkly among the studied cases of $N$. The \SU{2} case 
is a well-known and its ground state is N\'eel ordered. The correlations are expected to retain their sign structure from high temperatures down to $T/J=0$, while the structure
factor remains peaked at $(\pi,\pi)$ for all $T/J<\infty$.

The \SU{3} Heisenberg model on the square lattice also shows long-range spin order~\cite{Toth2010,Bauer2012}, with an ordering 
wave vector $\pm (2\pi/3,2\pi/3)$ or $\pm (-2\pi/3,2\pi/3)$. The two distinct orientations differ in their sign of the spin correlations across the diagonal of a square plaquette. 
This difference can be elevated to an Ising order parameter which could order at finite temperature due to its discrete nature, despite the true long range order of the spin correlations being inhibited at 
finite temperature due to the Hohenberg-Mermin-Wagner theorem~\cite{Mermin1966,Hohenberg1967}. This scenario is similar to those put forward for frustrated \SU{2} 
systems~\cite{Chandra1990,Weber2003}, and being actively discussed in the context of nematic ordering in the pnictide superconductor materials~\cite{Fang2008,Xu2008}.

The ground state of the \SU{4} square lattice Heisenberg model is predicted to exhibit an even more involved spatial pattern of \SU{4} symmetry breaking~\cite{Corboz2011}. Here it is
also conceivable that the dimerization pattern orders at finite temperature, before true long-range order for the spins occurs at $T=0$.

In both the \SU{3} and the \SU{4} cases predicted the low-temperature regime is distinct from the Manhattan regime expected at high temperature. We discuss in the following complete ED simulations
on a $N_s=18$ square cluster for \SU{3} and a $N_s=4\times4=16$ sites cluster for \SU{4} and study the behaviour of real-space correlations and structure factors as a function of temperature $T/J$.
These complete numerical diagonalizations have been performed by adopting the \SU{N} Young tableaux basis~\cite{Nataf2014}, combined with large-scale SCALAPACK~\cite{SCALAPACK2019} highly parallel diagonalization
routines. The largest block to be diagonalized numerically had a dimension of almost 800'000. A microcanonical view on the \SU{3} data is discussed in App.~\ref{app:microcanonical}.

We start the discussion of the results for the real-space correlations of the two cases in Fig.~\ref{fig_su3_su4_correl}(a) and (b) for \SU{3} and \SU{4} respectively. The lines including the circle
symbols display the ED results evaluated at the corresponding temperatures. The dashed lines on the other hand display the leading order behaviour Eq.~\eqref{eq:highT_spincorrs} of the Manhattan regime.
The high-temperature sign structure in the ED data in both cases is in perfect agreement with our Manhattan regime prediction and extends down to an intermediate temperature 
of $T_D/J \approx 0.325$ for SU($3$) where the distance-$(3,0)$ correlator turns from negative to positive. For SU($4$) the distance-$(2,2)$ correlator
changes sign at $T_D/J\approx 0.575$, which is almost twice as high as in the $N=3$ case. In both cases the deviation from the Manhattan picture occurs at the largest possible
distance on the considered cluster, which is analogous to the one dimensional chain discussed above. It is noteworthy that the correlations beyond the nearest neighbor
distance remain quite small in the $N=4$ case compared to the $N=3$ case, even down to rather low temperatures.

In Fig.~\ref{fig_su3_su4_correl}(c) and (d) we show the structure factor for various distinct wave vectors as a function of $T/J$ for \SU{3} and \SU{4} respectively. As expected
we observe a maximum at $(\pi,\pi)$ in the high temperature regime. At low temperature we recognize a direct shift of the peak to the location expected in the ground state, 
i.e.~$\mathbf{k}=(2\pi/3,2\pi/3)$ and symmetry related momenta for $N=3$ and $\mathbf{k}=(\pi,\pi/2)$ and symmetry related momenta for $N=4$. These finite-size transitions occur at 
$T_L(N_s{=}18)/J\approx 0.15$ for $N=3$ and $T_L(N_s{=}16)/J\approx0.21$ for $N=4$. In the absence of larger systems allowing a finite-size analysis it remains open whether these
temperatures signal a first order phase transition from the short-range ordered Manhattan regime to the spatial symmetry broken low-temperature regime, or whether these 
features are indicators of Lifshitz temperatures separating two short-range ordered regimes, while a distinct symmetry breaking transition occurs at even lower temperature.
In case a Lifshitz temperature occurs first coming from high temperature, we can estimate the infinite system Lifshitz temperatures using the analysis in App.~\ref{app:SofKAnsatz}. We then obtain
$T_L(\infty)/J=0.180(5)$ for \SU{3} and  $T_L(\infty)/J=0.230(5)$ for \SU{4}. 

So in conclusion of this study of the \SU{3} and \SU{4} square lattice cases we can again confirm that the Manhattan regime
indeed accounts for the structure of spin correlations in real and momentum space from infinite temperature down to quite low temperatures. As in the 
chain case we observe that the real-space correlations signal a sign change at a higher temperature than the putative temperature of the Lifshitz transition
governing the structure factor.

\subsection{Triangular lattice: $N=3$}

As the last example we display the real space spin correlations as a function of temperature $T/J$ for the \SU{3} triangular lattice 
Heisenberg model in Fig.~\ref{fig:2d_triangular_su3_correlations} for $N_s=12$. This system size is not very large, it is however the largest we can diagonalize completely
while being compatible with the expected three sublattice ordered ground state physics~\cite{Lauchli2006,Tsunetsugu2006,Bauer2012}. At high temperature we expect the Manhattan regime to manifest itself, and indeed in
the left panel of Fig.~\ref{fig:2d_triangular_su3_correlations} we can see that the nearest neighbor correlation is negative, while the distance $\sqrt{3}$ and $2$ correlators are positive as they both belong to the 
$m=2$ Manhattan shell. A similar Manhattan structure of correlators on the triangular lattice has recently been observed at short times in a non-equilibrium Rydberg quantum magnetism experiment~\cite{Lienhard2018}. At a disorder temperature $T_D(N_s{=}12)/J\approx 0.69$ the distance 2 correlator changes sign, then reaching the expected sign structure of the three sublattice ordered ground state. This disorder temperature is about two times
larger than in the square lattice case. It remains to be seen whether this is a finite-size effect or due to the different geometry. 

In the right panel of Fig.~\ref{fig:2d_triangular_su3_correlations} we present the corresponding structure factor $\mathcal{S}(T,\mathbf{k})$ as a function of temperature $T/J$. On the triangular lattice the Manhattan regime leads to a (shallow) peak at
the $K$ points in the Brillouin zone for all $N$, and that is well reproduced in our data. For $N=3$ the peak remains at the $K$ momenta for all temperatures, as the ground state develops Bragg peaks at these very momenta.
So there is no Lifshitz temperature for the triangular lattice $N=3$, despite the fact that some of the correlations change their sign in real space as a function of temperature, signalling the breakdown of the Manhattan regime.

\begin{figure}[!htb]
\centering
\includegraphics[width=0.48\textwidth]{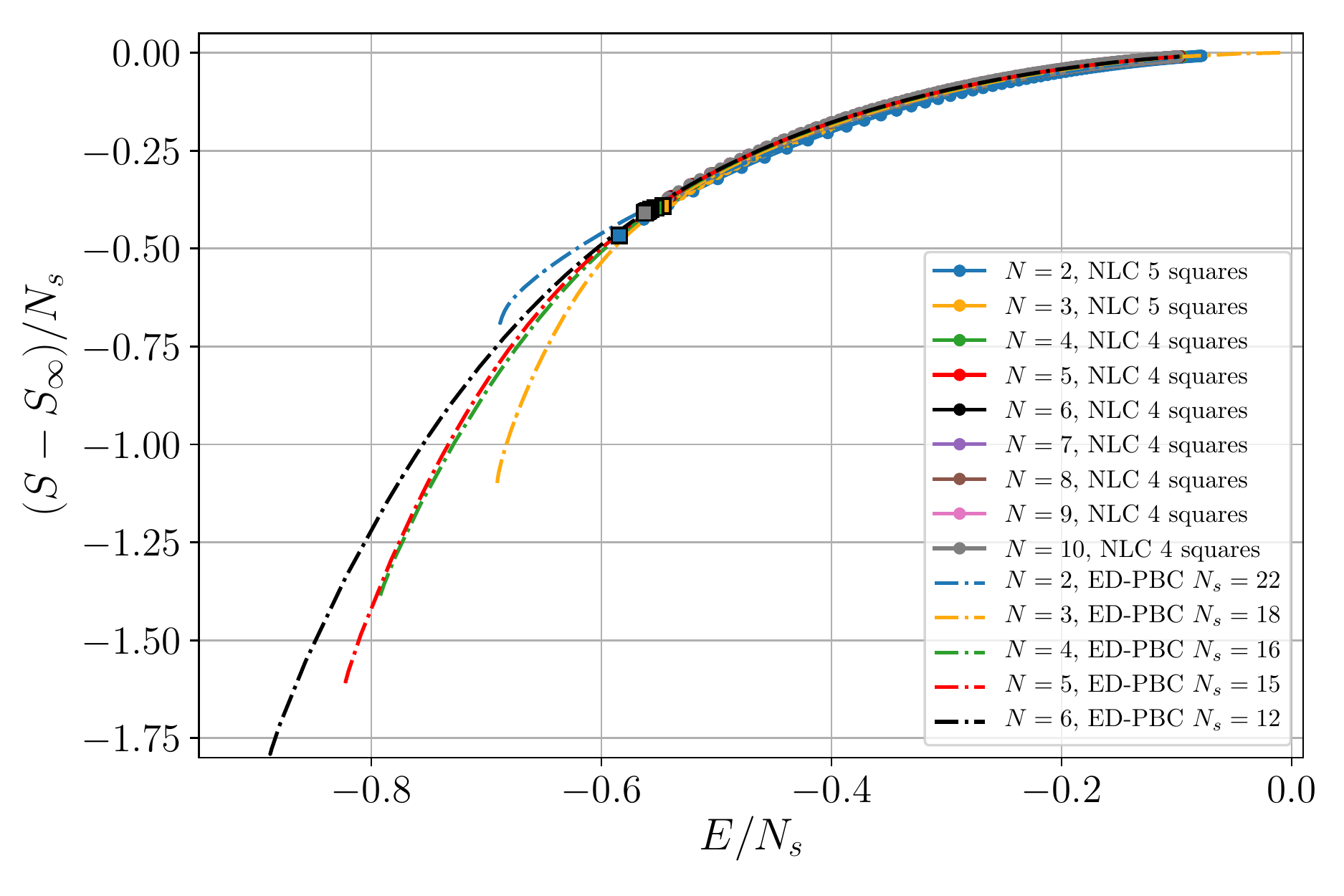}
\caption{Entropy per site as a function of the energy per site for the \SU{N} Heisenberg model
on the square lattice. The full lines with the filled circles denote NLC results, while the dot-dashed lines
denote complete numerical diagonalizations for periodic finite size clusters. The circles are plotted at the 
$(s,e)$ coordinate corresponding to $T/J=1/2$ for all $N$.}
\label{fig_highT_entropy_energy}
\end{figure}

\section{Equation of state on the square lattice}
\label{sec:eqationofstate}

In the discussion so far we used only the temperature $T/J$ as a control parameter of the thermal equilibrium. However in the
ultracold atom context it is also useful to understand the physics in terms of the entropy $S$ or the entropy per site $s$.
In order to address the entropy dependence of the correlations (as e.g.~studied in Ref.~\cite{Messio2012} for one dimensional \SU{N} chains)
we numerically determine the entropy per site $s\equiv S/N_s$ as a function of the energy density $e\equiv E/N_s$. 
This function $s(e)$ is known to be a thermodynamic potential, and allows therefore to extract e.g.~the temperature via $$\frac{ds}{de}=\frac{1}{T(e)}\ .$$ 
We note that for our nearest neighbour spin Hamiltonians \eqref{eq:Hspin} the energy per site $e=\frac{z}{2}\ C_\mathrm{spin}(n.n.)$ is related to the nearest neighbor spin correlator discussed above
via the coordination number $z$ of the lattice.

We calculated the entropy $S(T)$ and the energy $E(T)$ from the finite size partition function obtained by complete numerical diagonalizations of periodic square lattice systems. Furthermore we have
implemented a numerical linked cluster expansion based on a real space cluster expansion in terms of squares~\cite{Rigol2007a}, while the complete
numerical diagonalizations required for each cluster were performed in the \SU{N} Young tableau basis~\cite{Nataf2014}.

In Fig.~\ref{fig_highT_entropy_energy} we show the resulting curves $s(e)$ for various $N$ between $2$ and $10$ for the square lattice. We chose the origin
of the $y$-axis at the infinite temperature reference value $\ln(N)$. We have indicated the $(s,e)$ coordinates corresponding to $T/J=1/2$ with filled squares 
in Fig.~\ref{fig_highT_entropy_energy}. Curiously we observe that almost all curves lie on top of each other in the regime corresponding
to high temperature. The consequence of this observation is that in the high temperature regime of \SU{N} Heisenberg models one has to shelve away an $N$-{\em independent}
amount of entropy per site to cool to the same temperature (here $T/J=1/2$ as an example). Obviously this is not true when cooling to the ground state, as then the full entropy
per site of $\ln(N)$ has to be removed.

\begin{figure}[t]
\centering
\includegraphics[width=0.48\textwidth]{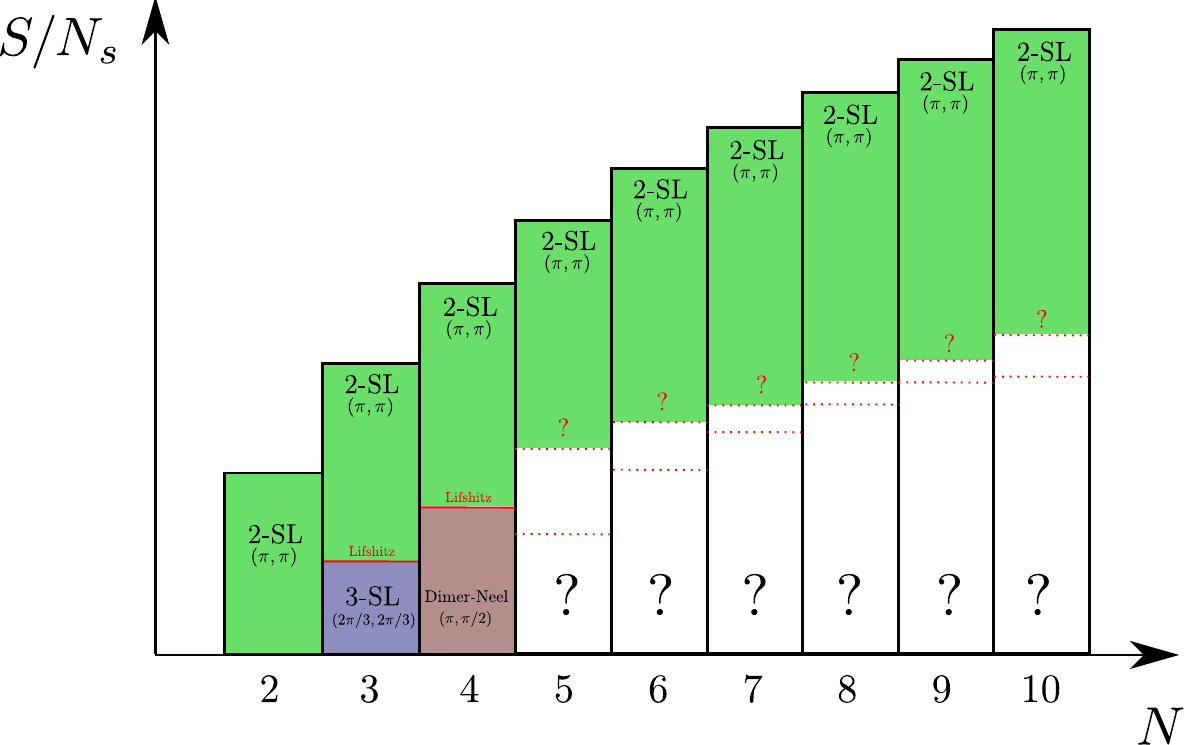}
\caption{Schematic phase diagram of the SU($N$) Heisenberg model on the square lattice for $N\le 10$. The maximum achievable entropy per site for a given $N$ is $\log N$ 
(corresponding to $T\rightarrow \infty$) signalized by the height of the bars. }
\label{fig_sun_square_phasediagram}
\end{figure}

Our leading order high temperature series expansion result \eqref{eq:highT_spincorrs} allows us to understand the origin of this phenomenon. In this expansion 
the nearest neighbour correlations ($m=1$) depend on $N$ as $(1-1/N^2)$, becoming basically $N$-independent rather quickly. On the one hand this correlator is proportional
to the energy $e$, and on the other hand we can determine the entropy reduction away from infinite temperature from an integration of the specific heat per site
$c(T)=de/dT$:
$$ \Delta s(T)= \ln(N) - s(T) = \int_T^{\infty} \frac{c(\tau)}{\tau} d\tau$$
We thus see that the approximate $N$-independence of the nearest-neighbor correlator in the Manhattan regime, and its relevance for the energy, implies that the entropy reduction to
reach a certain final temperature is approximately $N$-independent on the type of lattices considered in this work. 

This also explains the two apparently conflicting results regarding one-dimensional chains, that the Lifshitz temperatures are almost independent of $N$ according to Ref.~\cite{Fukushima2002}, while Ref.~\cite{Messio2012}
reports an increasing entropy per site for the Lifshitz points. Here we see that the two point of views coincide when viewed from infinite temperature, but seem to diverge when viewed from zero
temperature.

Unfortunately our methods do not allow us to systematically and reliably study the low temperature physics of large $N$ and $N_s$ systems, so a large
fraction of the $s$-$N$ phase diagram for the square lattice in Fig.~\ref{fig_sun_square_phasediagram} has to remain uncharted. However our work has substantiated
an extended region in temperature or entropy where the correlations are short-ranged and structured in real-space according to the Manhattan structure. This region
is indicated in green in Fig.~\ref{fig_sun_square_phasediagram}. For $N=2$ this Manhattan region is continuously connected to the low temperature region with an exponentially
diverging correlation length for the same structure of correlations. For $N=3$ and $N=4$ we have worked out the approximate location of the Lifshitz or first order transitions to a different
low-temperature regime in Sec.~\ref{subsec:square_N_3_4}. For larger $N$ we are unable to reliably estimate the lower end of the Manhattan region.

\section{Conclusion}

In this work we have analyzed the real-space structure of spin correlations in the \SU{N} Heisenberg model with spins
in the fundamental representation on a broad range of lattices. We find a unifying pattern, the Manhattan structure, where
spin correlations are organized in shells of equal Manhattan distance, and alternating in sign from one shell to the next. 

For selected case we have investigated how the Manhattan regime breaks down at low temperature through indicators such
as the disorder or Lifshitz temperature. 

Investigating the dependence of the entropy reduction from the infinite temperature value of $\ln(N)$, we have realized that 
the Manhattan regime is governed by an approximately $N$-independent equation of state, see Fig.~\ref{fig_highT_entropy_energy}.
This has interesting consequences, such that the entropy reduction from $\ln(N)$ to reach a certain temperature in the Manhattann regime
is approximately $N$ independent, potentially easing the way to reach low temperatures in the \SU{N} spin models, akin to the Pomeranchuk 
cooling effect discussed previously~\cite{Hazzard2012,Taie2012,Cai2013}.

An important open question remains however. While reaching low temperatures in the Manhattan regime seems easy, it is not clear how easy 
it will be to go to even lower temperatures where more $N$ specific novel physics can be reached. For example in the one-dimensional chains the
\SU{N} Wess-Zumino-Witten regime predicted at zero temperature is visible only at temperatures which decrease with $N$, as discussed 
in Ref.~\cite{Fukushima2002}. The temperatures we report to reach the ground state physics of the \SU{3} and \SU{4} square lattices with $T_L/J\approx 0.18$
and $T_L/J\approx 0.23$ are already quite low.

\subsection*{Acknowledgments}
AML thanks S. F\"olling for discussions which motivated the present investigation. 
We acknowledge support by the Austrian Science Fund (project IDs: F-4018 and I-2868). 
The computational results presented have been achieved in part using the Vienna Scientific Cluster (VSC).
This work was supported by the Austrian Ministry of Science BMWF as part of the UniInfrastrukturprogramm 
of the Focal Point Scientific Computing at the University of Innsbruck. We acknowledge PRACE for granting access to 
  "Joliot Curie" HPC resources at TGCC/CEA under grant number 2019204846.

\appendix
\section{Observables for \SU{N} spin correlations}
\label{app:spincorrs}
In this section we discuss the relation between several observables which
are useful to quantify spin correlations for \SU{N} quantum spin systems with
local spins in the $N$-dimensional fundamental ( ${\tiny \yng(1)}$ ) irreducible
representation of \SU{N}. Assuming an arbitrary state $\rho$ of the entire system which is
\SU{N} invariant (e.g.~an \SU{N} singlet pure state, or a non-symmetry-broken 
thermal density matrix), the two site reduced density matrix $\rho_{i,j}=\Tr_{E(i,j)}\left[\rho\right]$,
(where $E(i,j)$ denotes all remaining degrees of freedom apart from sites $i$ and $j$) contains all the
information regarding correlations between sites $i$ and $j$.
The two site reduced density matrix of linear dimension $N^2$ has two subspaces:
the symmetric subspace (${\tiny \yng(2)}$) of dimension $N(N+1)/2$ and the antisymmetric
(${\tiny \yng(1,1)}$) of dimension $N(N-1)/2$. The total weight of the state on the symmetric and the
antisymmetric subspaces are denoted $p_S$ and $p_A$ respectively, with $p_S+p_A=1$.

Let us now discuss a few observables and their relation to $p_S$ and $p_A$. We use the notation
$\langle \mathcal{O}_{i,j}\rangle=\Tr\left[\mathcal{O}_{i,j} \rho_{i,j}\right]$.
\begin{enumerate}
\item The two site permutation operator $P_{i j}$ is particularly simple in this respect. The 
operator has eigenvalue $+1$ $(-1)$ in the symmetric (antisymmetric) subspace. 
\begin{equation}
C_P(i,j)\equiv\langle P_{i,j}\rangle= p_S-p_A
\end{equation}
At infinite temperature, the two-site reduced density matrix is proportional to the identity: $\rho_{i,j}(T{=}\infty)=\mathbb{1}/{N^2} $. This leads to $p_S=\frac{N(N+1)}{2N^2}$ and
$p_A=\frac{N(N-1)}{2N^2}$. Thus the correlator:
$C_P(i,j)=1/N$ at $T=\infty.$
\item The contraction of \SU{N} spin operators which we use in the Hamiltonian Eq.~\eqref{eq:Hspin} are related to the permutation operator as
outlined in Eq.~\eqref{eq:Hspin_perm}. This leads to the following correlations:
\begin{eqnarray}
  C_\mathrm{spin}(i,j)&=& \langle \sum_{A} S^{A}_{i}  S^{A}_{j}\rangle\\
               &=& \tfrac{1}{2} ( C_P(i,j)-\tfrac{1}{N} )\nonumber\\
               &=& \tfrac{1}{2} (p_S-p_A-1/N)\nonumber
\label{eq:SdotScorr}
\end{eqnarray}
This operator reduces to the well-known $\mathbf{S}_i\cdot \mathbf{S}_j$ operator for $S{=}1/2$ for \SU{2}, yielding $C_\mathrm{spin}(i,j)=-3/4\ (+1/4)$ for an \SU{2} singlet (triplet).

At infinite temperature this correlation vanishes for all $N$:
$C_\mathrm{spin}(i,j)=0$ at $T=\infty$.

\item In the recent "Singlet-Triplet-Oscillations" (STO) experiments for \SU{2} and \SU{N{>}2} systems~\cite{Greif2013,Ozawa2018,TakahashiICAP2018,TakahashiDAMOP2020} 
it is possible to estimate the symmetric ($p_S$, "triplet") and the antisymmetric ($p_A$, "singlet") fraction of a nearest neighbour density matrix on several lattices (e.g.~honeycomb, 
cubic). We refer to those
references for the details of the method. 

\item We anticipate that in future quantum gas microscopes it will be possible to record snapshots of the internal spin state
configurations of a cloud of atoms, in analogy to what is currently possible for \SU{2} fermions in an optical 
lattice~\cite{Parsons2016,Boll2016,Cheuk2016,Brown2017,Mazurenko2017}. In such experiments it is then possible to 
measure "color-color" correlations, i.e.~to measure the probability of finding two atoms at sites $i$ and $j$ in the
same internal \SU{N} spin state $\alpha$.

We therefore define a diagonal color correlator for \SU{N} spin models as
\begin{eqnarray}
 C_\mathrm{color}(i,j)  &=& \langle \sum_{\alpha}  |\alpha_i,\alpha_j\rangle\langle \alpha_i ,\alpha_j| \rangle \nonumber\\
     &=& \langle \sum_{\{ A \in C \}} 2 S^{A}_{i} S^{A}_{j} + \tfrac{1}{N} \rangle 
                  \label{eq:CcolorCorr}
\end{eqnarray}
where $C$ denotes the set of $N-1$ indices corresponding to the diagonal spin operators $S^{A}_{i}$, $S^{A}_{j}$ on 
site $i$ and $j$, i.e. the hermitian Cartan generators of the Lie algebra $su(N)$.
This observable is a projector and can be seen as the probability to have the same spin color on
        site $i$ and $j$. For example, in the case of \SU{2} and \SU{3} the observable reads 
$C_\mathrm{color}(i,j)=2 \langle S^z_{i} S^z_{j}\rangle + 1/2$ and $C_\mathrm{color}(i,j) = 2 (\langle S^3_i S^3_j\rangle + \langle S^8_i S^8_j\rangle ) + 1/3$, respectively.

Due to the \SU{N} symmetry the expectation value $\langle S^A_i S^A_j \rangle$ is the same for every component $A$ and hence
$\langle \sum_{A} S^A_i S^A_j \rangle = (N^2-1) \langle S^0_i S^0_j \rangle$.
After some algebraic steps one obtains the relation of the diagonal color correlator with the 
two site permutation operator 
\begin{eqnarray}
    C_\mathrm{color}(i,j) &=& \tfrac{1}{N} + \tfrac{N-1}{N^2-1} (\langle P_{ij} \rangle - \tfrac{1}{N})\nonumber\\
    	&=& \tfrac{1}{N} + \tfrac{1}{N+1}  (p_S-p_A - \tfrac{1}{N}) 
\end{eqnarray}
At infinite temperature the $C_\mathrm{color}(i,j)$ correlator takes the value $1/N$.

\end{enumerate}

\section{Microcanonical Analysis}
\label{app:microcanonical}
\begin{figure*}
\centering
\includegraphics[width=\linewidth]{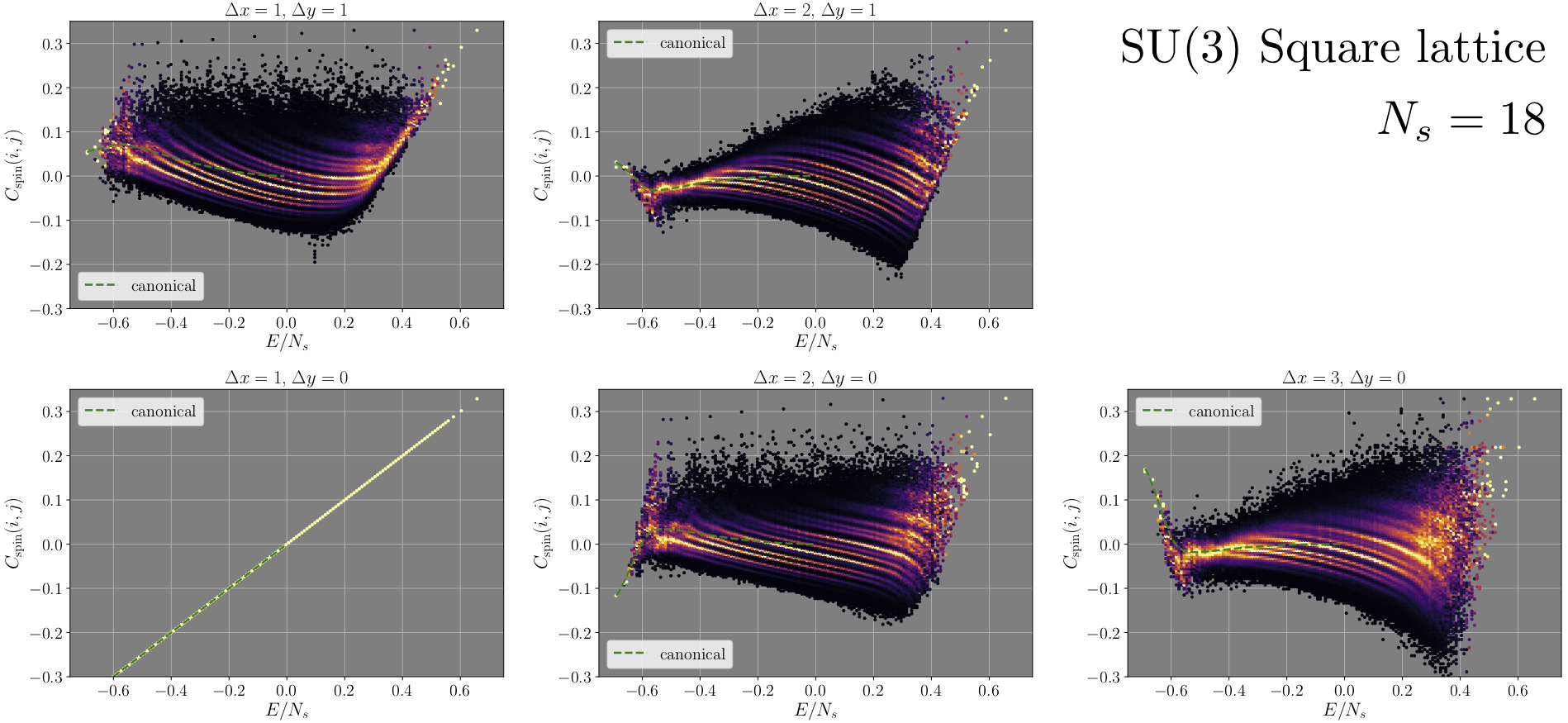}
\caption{Microcanonical analysis of spin correlations for the \SU{3} Heisenberg model on a 18 sites square lattice cluster, For a detailed description see App.~\ref{app:microcanonical}.}
\label{fig:overview_microcanonical_SU3_square}
\end{figure*}

The material in the main text is derived from the canonical Gibbs ensemble. In this appendix we
highlight selected results for spin correlations on square lattice clusters, where large-scale numerical full diagonalizations
exploiting the \SU{N} symmetry~\cite{Nataf2014} have been carried out. In Fig.~\ref{fig:overview_microcanonical_SU3_square}
we show results for a spatially symmetric 18 sites square cluster for the \SU{3} Heisenberg model. This cluster has periodic boundary conditions and is spanned by the 
simulation cell vectors $\mathbf{T}_1=(3,3)$ and $\mathbf{T}_2=(-3,3)$. The entire Hilbert space has a dimension of $3^{18}=387\ 420\ 489$.
After dividing the Hilbert space into \SU{N} irreducible representations, the largest matrix size to be diagonalized features a dimension of almost $800\ 000$.
The panels are arranged according to the position of the second site in the correlator with the origin. In each panel we show a two-dimensional histogram compiled from all 
$3^{18}$ eigenstates of the system, where on the $x$ axis the energy per site $E/N_s$ is plotted, while on the $y$ axis the value of the corresponding correlator $C_\mathrm{spin}(\Delta x,\Delta y)$.
Furthermore we plot the energy-correlator behaviour of the canonical predictions in the temperature range $T=0$ to $T=\infty$, based on the same finite size data (dark green, dashed line). The positive energy 
per site region corresponds to the ferromagnetic side of the energy spectrum. One can see how the ferromagnetic correlations build up as one approaches the maximum energy per site ($E/N_s=2/3$). 
On the antiferromagnetic side of the energy spectrum the behaviour of the correlators is less regular, but one can recognize that at low energy, the distribution of the correlators starts to concentrate and converge towards the $T=0$, 
i.e.~ground state results, where the canonical and the microcanonical predictions match.

Overall one can see that the expected eigenstate thermalization hypothesis (ETH)~\cite{Deutsch1991,Srednicki1994,Beugeling2014} behaviour is not fully reached yet for this system size, 
despite the huge total Hilbert space. We attribute this to the large amount of different quantum number sectors (spatial symmetry sectors combined with \SU{3} representations)
which contribute to the observables, and which are all included in the plots.

\section{Continuous structure factor ansatz}
\label{app:SofKAnsatz}
In this section we perform an continuous structure factor ansatz and thereby estimate the Lifshitz temperature under the assumption that coming from high 
temperatures a Lifshitz
transition occurs. 
We focus on the 2D square lattice but for the one-dimensional chain the derivation is perfomed in an analog way.
We start by decomposing the static structure factor of the spin-spin correlations on the infinite square lattice into harmonics
\begin{equation}
 S(\mathbf{k}) = {\sum_{j\neq 1}} ' 2 \cos[\mathbf{k}(\mathbf{r}_j-\mathbf{r}_0)] \left< \mathbf{S}_j \mathbf{S}_1 \right> + \frac{N^2-1}{2N},
\label{eq_strucfac_ansatz}
\end{equation}

where the sum ${\sum}'$ runs over the subset corresponding to the indices of one quadrant of the lattice, centered by the reference site $\mathbf{r}_0$. 
We cut the series at distance three and set correlations for larger distances to zero by definition
\begin{eqnarray}
    S(\mathbf{k})  &=&c_0 + c_1 [\cos(k_x) + \cos (k_y)] \nonumber \\
                 &+&  c_2 [\cos (2k_x) + \cos (2k_y)] \nonumber \\
                 &+& c_3 [\cos (k_x+k_y)+\cos(k_x-k_y)] \nonumber \\
                 &+& c_4 [\cos (3k_x) + \cos (3 k_y)] \nonumber \\
                 &+& c_5 [\cos (2k_x+k_y) + \cos (2k_x-k_y) \nonumber \\
                 &+& \cos (k_x+2k_y) + \cos (k_x-2k_y)],
\end{eqnarray}
where $c_j = 2 \, C_{\mathrm{spin}}(0,j) $.
Taylor expanding around $(\pi,\pi)$ and keeping maximally quartic terms leads to the following Ginzburg-Landau free energy like expression
\begin{equation}
 S((\pi+x,\pi+y)) \approx \; a_0 + a_1(x^2+y^2)+a_2 (x^4+y^4)+a_3 x^2 y^2
\label{app:eqGinzburg}
\end{equation}
with
\begin{equation}
 \begin{split}
  a_0 &= c_0-2c_1+2c_2+2c_3-2c_4-4c_5\\
  a_1 &= \frac{c_1}{2}-2c_2-c_3+\frac{9c_4}{2}+5c_5\\
  a_2 &= -\frac{c_1}{24}+\frac{2c_2}{3}+\frac{c_3}{12}-\frac{27c_4}{8}-\frac{17c_5}{12}\\
  a_3 &= \frac{c_3}{2}-4c_5.
 \end{split}
\end{equation}
By maximizing Eq.~\eqref{app:eqGinzburg} we find three different regimes: (I) a trivial regime with the maximum at $(\pi,\pi)$, (II) a second regime
with four maxima along the diagonals ($(\pi\pm\epsilon,\pi\pm\epsilon)$ with $\epsilon=\sqrt{-a_1/(2a_2+a_3)}$) through $(\pi,\pi)$  and (III) a
 third regime with four maxima along horizontal and vertical lines ($(0,\pi\pm \epsilon)$ and $(\pi\pm\epsilon,0)$ with $\epsilon=\sqrt{-a_1/(2a_2)}$) through $(\pi,\pi)$.
Finally, by matching the coefficients ${a_i}$ with the ED data for our largest cluster for every temperature we obtain a trajectory in the coefficient space.
Starting from the origin ($T\rightarrow\infty$) 
the trajectory moves into the trivial regime, where the structure factor is peaked at $(\pi,\pi)$, but bends back and breaks through regime (II) 
at $T_L(\infty)/J=0.180(5)$ for \SU{3} and through regime (III) at $T_L(\infty)/J=0.230(5)$ for \SU{4}, which are the estimated Lifshitz temperatures.

\bibliographystyle{apsrev4-1_custom}
\bibliography{references_SU_N_thermo}

\end{document}